\pdfoutput=1
\documentclass[12pt]{article}
\usepackage{jheppub}
\usepackage[utf8]{inputenc}
\usepackage{lmodern, microtype}
\usepackage{amsmath}
\usepackage{amssymb}
\usepackage{graphicx,color}
\usepackage{amsthm,booktabs}
\usepackage{mathrsfs,blkarray,dsfont}
\usepackage{tikz}
\usetikzlibrary{matrix,arrows}

\usepackage{ifpdf}
\usepackage{breqn,nicefrac}
\usepackage{subfigure}

\usepackage[boxsize=0.5em,aligntableaux=center]{ytableau}

\usepackage{tikz}
\usetikzlibrary{shapes,shadows,calc}
\usepgflibrary{arrows}
\usepackage{rotating}
\usepackage{wrapfig}

\DeclareFontFamily{U}{mathx}{\hyphenchar\font45}
\DeclareFontShape{U}{mathx}{m}{n}{<-> mathx10}{}
\DeclareSymbolFont{mathx}{U}{mathx}{m}{n}
\DeclareMathAccent{\widebar}{0}{mathx}{"73}

\newcommand{\mfrac}[2]{\tfrac{#1}{\rule{0pt}{0.58em} #2} \,}
\newcommand{\bA}{\mathbf{A}}
\newcommand{\bB}{\mathbf{B}}

\newcommand{\bI}{\mathbf{I}}
\newcommand{\bJ}{\mathbf{J}}

\newcommand{\dm}[1]{{\dd\,}^{#1}}
\newcommand{\wb}{\widebar}

\newcommand{\hK}{\mathcal{K}^{\mathbf b}}

\DeclareMathOperator{\re}{Re}

 \makeatletter
 \def\BorderMatrix#1{%
 	\begingroup
 	\m@th\@tempdima8.75\p@\setbox\z@\vbox{%
 		\def\cr{\crcr\noalign{\kern2\p@\global\let\cr\endline }}%
 		\ialign{\hfil$##$\hfil\kern2\p@\kern\@tempdima & \thinspace 
 			\hfil $##$\hfil && \quad\hfil $##$\hfil\crcr\omit\strut
 			\hfil\crcr\noalign{\kern-\baselineskip}#1\crcr\omit
 			\strut\cr}}%
 	\setbox\tw@\vbox{\unvcopy\z@\global\setbox\@ne\lastbox}%
 	\setbox\tw@\hbox{\unhbox\@ne\unskip\global\setbox\@ne\lastbox}%
 	\setbox\tw@\hbox{%
 		$\kern\wd\@ne\kern-\@tempdima\left(\kern2pt
 		\global\setbox\@ne\vbox{\box\@ne\kern2\p@}%
 		\vcenter{\kern -\ht\@ne \unvbox\z@\kern-\baselineskip}%
 		\kern-2\@tempdima\kern4\p@\right)$}%
 	\null\;\vbox{\kern\ht\@ne\box\tw@}%
 	\endgroup
 }
 \makeatother

\tikzset{
  sshadow/.style={opacity=.25, shadow xshift=0.05, shadow yshift=-0.06},
}

\def\kbbox[#1,#2,#3,#4,#5]#6{
        \draw[dashed] node[draw,color=gray!50,minimum
        height=#1,minimum width=#2] (#4) at #5 {}; 
        \node[anchor=#3,inner sep=2pt] at (#4.#3)  {#6};
}
\def\kbboxred[#1,#2,#3,#4,#5]#6{
        \draw[] node[draw,color=red,minimum
        height=#1,minimum width=#2] (#4) at #5 {}; 
        \node[anchor=#3,inner sep=2pt] at (#4.#3)  {#6};
}

\newcommand{\cA}{\mathcal{A}}
\newcommand{\cB}{\mathcal{B}}
\newcommand{\cC}{\mathcal{C}}
\newcommand{\cD}{\mathcal{D}}

\newcommand{\cF}{\mathscr{F}}

\newcommand{\cI}{\mathcal{I}}
\newcommand{\cJ}{\mathcal{J}}
\newcommand{\cK}{\mathcal{K}}
\newcommand{\cL}{\mathcal{L}}

\newcommand{\cN}{\mathcal{N}}

\newcommand{\cV}{\mathcal{V}}

\newcommand{\bbZ}{\mathbb{Z}}

\newcommand{\bbC}{\mathbb{C}}

\newcommand{\wt}[1]{\widetilde{#1}}

\DeclareMathOperator{\im}{Im}

\def\a{{\alpha}}
\def\b{{\beta}}

\def\Lam{{\Lambda}}
\def\lam{{\lambda}}

\def\g{{\gamma}}
\def\i{{\iota}}
\def\k{{\kappa}}

\def\p{{\partial}}
\def\l{{\langle}}

\def\bs#1\es{\begin{split}#1\end{split}}
\def\ba#1\ea{\begin{align}#1\end{align}}
\def\baed#1\eaed{\begin{aligned}#1\end{aligned}}
\def\bged#1\eged{\begin{gathered}#1\end{gathered}}

\def\a{\alpha}
\def\b{\beta}

\def\d{\delta}

\def\g{\gamma}

\def\k{\kappa}
\def\l{\lambda}
\def\L{\Lambda}
\def\m{\mu}

\def\o{\omega}

\def\S{\Sigma}

\def\Re{\text{Re}}

\newcommand{\ws}{\wedge \star}
\newcommand{\rf}{\re f}
\newcommand{\w}{\wedge}

\def\we{\wedge}

\let\foo\bar 
\renewcommand{\bar}[1]{ {\foo{  #1} }{} }

\newlength{\dhatheight}

\def\ih{{\hat \i}}
\def\jh{{\hat \jmath}}
\DeclareMathOperator{\dd}{d \!}
\def\m{{m}}

\def\Label#1{\label{#1}%
  \smash{\hbox to0pt{\raise1ex\hbox{\tiny[#1]}\hss}}}
\def\noLabels{\let\Label=\label}
\def\nobbibitem{\let\bbibitem=\bibitem}
 \def\noBibitem{\let\Bibitem=\bibitem}

\newcommand{\be}{\begin{equation}}
\newcommand{\ee}{\end{equation}}
\newcommand{\beq}{\begin{equation}}
\newcommand{\eeq}{\end{equation}}
\newcommand{\bea}{\begin{eqnarray}}
\newcommand{\eea}{\end{eqnarray}}

\newcommand\varpm{\mathbin{\vcenter{\hbox{%
  \oalign{\hfil$\scriptstyle+$\hfil\cr
          \noalign{\kern-.3ex}
          $\scriptscriptstyle({-})$\cr}%
}}}}
\newcommand\varmp{\mathbin{\vcenter{\hbox{%
  \oalign{$\scriptstyle({+})$\cr
          \noalign{\kern-.3ex}
          \hfil$\scriptscriptstyle-$\hfil\cr}%
}}}}

\title{\centering \Large Shift-symmetries and  gauge coupling functions   \\
 in orientifolds and F-theory}

\author[]{Pierre Corvilain,}
\author[]{Thomas W.~Grimm}
\author[]{and Diego Regalado}

\affiliation[]{Max Planck Institute for Physics,\\
F\"ohringer Ring 6, 80805 Munich, Germany\\[.1cm]
\text{\normalfont and}}

\affiliation[]{
Institute for Theoretical Physics and \\
Center for Extreme Matter and Emergent Phenomena,\\
Utrecht University, Leuvenlaan 4, 3584 CE Utrecht, The Netherlands}

\emailAdd{pcorvila@mpp.mpg.de}
\emailAdd{grimm@mpp.mpg.de}
\emailAdd{regalado@mpp.mpg.de}

\abstract{We investigate the field dependence of the gauge coupling 
functions of four-dimensional Type IIB orientifold and F-theory compactifications 
with space-time filling seven-branes. In particular, we analyze the constraints 
imposed by holomorphicity and covariance under shift-symmetries of the bulk and brane 
axions. This requires introducing quantum corrections that necessarily contain Riemann theta functions 
on the complex torus spanned by the D7-brane Wilson line moduli. Our findings hint
towards a new underlying geometric structure for gauge coupling functions in string compactifications.
We generalize this discussion to a genuine F-theory compactification 
on an elliptically fibered Calabi-Yau fourfold. 
We perform the first general 
dimensional reduction of eleven-dimensional supergravity and dualization to 
the F-theory frame. The resulting  effective action 
is compared with the circle reduction of a four-dimensional $\cN=1$ supergravity theory. 
The F-theory geometry elegantly unifies bulk and brane degrees of freedom and 
allows us to infer non-trivial results about holomorphicity and shift-symmetries. 
For instance, we gain new insight into kinetic mixing of bulk and brane gauge fields.}

\begin{document}
\setlength{\parskip}{5pt}

\makeatletter
\renewcommand{\@fpheader}{
\hfill MPP-2016-148}
\makeatother

\maketitle
\newpage

\section{Introduction}

In four-dimensional effective actions with minimal $\cN=1$ supersymmetry, the dynamics of 
the vector fields crucially depends on the gauge coupling functions determining their kinetic 
terms. Supersymmetry requires this function to be holomorphic in the 
complex scalars that arise as the bosonic parts of chiral multiplets~\cite{Wess:1992cp}. 
This holomorphicity allows to infer certain non-renormalization theorems 
for this coupling function. In particular, one can show that it only 
receives perturbative corrections to one-loop order, while
non-perturbative corrections can be generally present.
In effective theories arising from string theory, the 
gauge coupling function can depend on scalars admitting 
classical shift-symmetries. While this is key in the implementation of
anomaly cancellation via the Green-Schwarz mechanism \cite{Green:1984sg,Sagnotti:1992qw}, these 
symmetries can also constrain the functional form of the coupling independent 
of any gauging. In this work we exploit the interplay between holomorphicity 
and symmetries in the study of gauge coupling functions of brane and 
R-R gauge fields.

Deriving the gauge coupling function in a full-fledged string model can be challenging.
In intersecting D-brane models this function has been investigated since 
their first construction~\cite{Blumenhagen:2006ci,Ibanez:2012zz}. 
Of particular interest in this work will be intersecting Type IIB D-brane models 
with space-time filling D7-branes and O7-planes and their generalizations 
to F-theory models with seven-branes of general type. We 
furthermore focus on compactifications yielding a four-dimensional 
effective theory with $\cN=1$ supersymmetry.
At weak string coupling, i.e.~when D7-branes and O7-planes are considered,
the gauge coupling function can be studied by dimensionally reducing the 
D7-brane effective action as done in~\cite{Jockers:2004yj,Jockers:2005zy}.   

Interestingly, it was already pointed out in~\cite{Jockers:2004yj} (and for the mirror-dual
configurations in~\cite{Grimm:2011dx}) that the gauge coupling functions determined 
by direct classical reduction are not holomorphic in the complex coordinates determined 
for the rest of the effective action. First, this was observed for 
the D7-brane gauge coupling function in the presence of D7-brane Wilson line 
moduli. A solution to this problem was, however, suggested in~\cite{Jockers:2004yj},
by arguing that the missing terms arise at one-string-loop order by using 
the orbifold results of~\cite{Berg:2004ek,Berg:2004sj}. Second, including the mixing with R-R bulk U(1)'s, a further 
seeming conflict with holomorphicity in the independently derived complex 
coordinates is encountered. 
Given these gaps in our understanding of these basic couplings, one 
might wonder if there is a more systematic approach to determine and 
analyze these couplings. In this paper we suggest that by carefully studying
the shift-symmetries of the axions in the theory, one can significantly constrain the gauge coupling
function of both closed and open string gauge fields. This is done for the Type IIB weak string coupling setting in detail  
in section~\ref{sec_TypeIIB}, while the generalization to F-theory can be 
found in section~\ref{Fthy} and section~\ref{sec:four}. We should note however, that the F-theory analysis
is not simply a generalization, but it is also useful in uncovering new interesting facts
about the Type IIB case.

In general, the gauge coupling functions $\hat f$ of D-branes 
depend on R-R form axions of the underlying supergravity theory. 
Since these forms admit shift-symmetries, they 
can be used to constrain the functional dependence 
of $\hat f$ on the R-R-form axions. Using holomorphicity, one 
is then lead to constraints on the dependence of $\hat f$ on the 
complex coordinates. Clearly, exploiting the symmetry properties 
for determining the gauge coupling function in string compactifications
is not new and has, for example, already been discussed intensively 
in heterotic models (for early works on this subject, see 
e.g.~\cite{Kaplunovsky:1994fg,Kaplunovsky:1995jw} and references therein).
However, one fact that has not been exploited systematically 
is that higher-degree R-R forms can transform non-trivially under the 
shift-transformations of lower-degree R-R or D-brane gauge transformations.
This is a direct consequence of having Chern-Simons terms in higher dimensions
which, as we will discuss in detail, translates into having non-Abelian shift-symmetries
among the axions in the lower-dimensional effective field theory.

Our strategy to constrain the corrections to the gauge coupling function 
is to combine our knowledge of the appropriate $\cN=1$ complex coordinates 
with the expected symmetry properties of the gauge coupling function. 
More precisely, we first note that the gauge coupling function $\hat f_{\rm D7}$ is
proportional to the K\"ahler coordinates $T_\alpha$ in the absence of R-R and NS-NS two-form scalars $G^a$
and D7-brane Wilson line scalars $a_p$. Including these fields, one finds corrections to $T_\alpha$ 
depending on $G^a, a_p$ as well as their complex conjugates $\bar G^a, \bar a_p$. 
We argue that once these moduli are included, the gauge coupling function cannot be simply
given by $T_\a$, since that would break the discrete shift symmetries. However, just by using
holomorphicity and such discrete symmetries, we can derive that the correction to
$\hat f_{\rm D7}$ is a holomorphic section of a certain line bundle over the complex torus spanned
by the axions. Finally, this fixes the form of the corrections, which consist of logarithms of
Riemann theta functions depending on the Wilson lines.

The improved understanding of D7-brane gauge coupling functions finds 
an elegant description when moving to F-theory models studied 
via M-theory. In the F-theory description, the seven-brane dynamics 
is encoded by the geometry of an elliptically fibered Calabi-Yau 
fourfold $Y_4$. In particular, the complex structure moduli, seven-brane positions, 
and the axio-dilaton reside in a joint moduli space: the 
moduli space of complex structure deformations 
of $Y_4$. We also have that the two-form scalars $G^a$, Wilson lines $a_p$ and 
R-R gauge fields are unified as arising from elements of the third cohomology 
of $Y_4$. In fact, they parameterize the complex torus $H^{2,1}(Y_4)/H^{3}(Y_4,\bbZ)$.
The gauge coupling function can then be determined via the 
duality to M-theory on the same fourfold by the following procedure: (1)
compactify a general four-dimensional $\cN=1$ theory on a circle, (2)
integrate out all massive modes in the three-dimensional Coulomb branch, (3)
compare the result with an M-theory compactification on a smooth Calabi-Yau 
fourfold. Using this procedure, the leading seven-brane gauge coupling function was
found in~\cite{Grimm:2010ks} and some first results on corrections to this result have 
been obtained using this duality in~\cite{Grimm:2012rg}. As for the Type IIB case, we
expect that in general the gauge coupling function depends on the scalars $G^a$ and Wilson
lines. As of now, however, the contribution from two-form scalars and Wilson lines
has not been obtained via an M-theory reduction. Thus, in this work we will perform an M-theory
reduction on a generic elliptically fibered Calabi-Yau fourfold keeping 
track of all fields including the two-form scalars, the Wilson lines and the R-R
gauge fields, thereby generalising the results in~\cite{Grimm:2010ks}. We also explain in
great detail the relevance of having an elliptically fibered space and the dualization procedure to bring
the effective action to the correct F-theory duality frame to compare with a four-dimensional theory.
Exploiting the shift-symmetries in the M-theory reduction and the F-theory frame we present a 
detailed discussion of the F-theory gauge coupling function. We extend the analysis of \cite{Grimm:2015ona}
and propose quantum corrections to ensure holomorphicity and shift-symmetry invariance.

This work is organized as follows. In section~\ref{sec_TypeIIB} we discuss the 
$\cN=1$ effective action of a Type IIB orientifold compactification with 
a space-time filling D7-brane. We introduce the complex coordinates and 
K\"ahler potential capturing the dynamics of a rigid D7-brane with Wilson line moduli. 
We then study the symmetries of the moduli space and their action on the gauge
coupling function, which allows us to derive certain constraints for $\hat f$. In section~\ref{Fthy},
we perform the dimensional reduction of M-theory on a generic smooth Calabi-Yau
fourfold and dualize to the correct F-theory duality frame. We carefully derive the shift-symmetries of the effective theory
and the effect of the dualization on them. In section~\ref{sec:four} we determine the gauge
coupling function by matching the M-theory reduction with a circle reduction of a four-dimenisonal
theory.  Finally, we discuss the constraints that
holomorphicity and gauge-invariance imposes on it. We leave a detailed discussion of
the dualization of three-dimensional action to appendix~\ref{ap:dual} and of the circle reduction of
a four-dimensional theory to appendix~\ref{ap:circle}.

\section{The D7-brane gauge coupling function and kinetic mixing} \label{sec_TypeIIB}

In this section we consider the four-dimensional effective action that arises from Calabi-Yau orientifold compactifications of Type IIB with D7-branes and 
O7-planes. In particular, we aim to determine 
the characteristic functions determining the standard $\cN=1$ supergravity with bosonic action \cite{Wess:1992cp}
\begin{equation}
\begin{aligned}\label{act4D}
S^{(4)} = \int \frac 12  \hat R \, \hat \star \, 1 - \hat K_{\bA \wb \bB} \dd \hat M^\bA \wedge \hat \star \dd \widebar{\hat M}^{\bar \bB} & 
- \frac 14 \re \hat f_{\bI \bJ} (\hat M) \hat F^\bI \wedge \hat \star \hat F^\bJ \\
&\hspace{1cm}- \frac 14 \im \hat f_{\bI \bJ} (\hat M ) \hat F^\bI \wedge \hat F^\bJ   \, ,
\end{aligned}
\end{equation}
where $\hat K_{\bA \wb \bB}$ are the second derivatives of a real K\"ahler potential $\hat K(\hat M, \widebar{\hat M})$ and $\hat f_{\bI \bJ}(\hat M)$ is the holomorphic gauge coupling function. We will denote four-dimensional quantities with a hat. 
The functions $\hat K$, $\hat f_{\bI \bJ}$ as well as the complex coordinates $\hat M^\bA$ are determined by reducing Type IIB supergravity coupled to the D7-brane and O7-plane world-volume actions 
following and extending \cite{Grimm:2004uq,Jockers:2004yj,Jockers:2005zy}. We will also discuss the shift-symmetries and certain quantum 
corrections of the effective theory. 

\subsection{Complex coordinates and the K\"ahler potential in Type IIB orientifolds}

The general form of the effective action for the bulk fields in such compactifications was 
determined in~\cite{Grimm:2004uq} by reducing Type IIB supergravity on a Calabi-Yau manifold $Y_3$, while 
also including the action of an holomorphic involution $\sigma: Y_3 \rightarrow Y_3$. The 
action of $\sigma^*$ on the cohomology groups splits them into eigenspaces 
$H^{p,q}(Y_3) = H_+^{p,q}(Y_3)\oplus H_-^{p,q}(Y_3)$. The 
basis used to span these cohomologies is listed in table~\ref{BasisForms}.
\begin{table}[h!]
	\begin{center}
		\begin{tabular}{cccl}
			\toprule
				cohomology group &\quad basis elements\quad &\quad \quad fields \quad \quad  &  index range\\
			\midrule
			\hspace*{1.2cm} $H^{1,1}_+(Y_3)$ & $\omega_\alpha$ &  $v^\alpha,\, C_2^\alpha$ & $\alpha= 1, \ldots ,h^{1,1}_+(Y_3) $ \\
			\hspace*{1.2cm} $H^{1,1}_-(Y_3)$ & $\omega_a$ & $b^a,\, c^a$& $a = 1, \ldots ,h^{1,1}_- (Y_3)$ \\
			\hspace*{1.2cm} $H^{2,2}_+(Y_3)$ & $\tilde \omega^\alpha$ &  $\rho_\alpha$&$\alpha = 1, \ldots ,h^{1,1}_+(Y_3) $ \\
			\hspace*{1.2cm} $H^3_+(Y_3)$ & $(\alpha_\kappa, \beta^\kappa)$ & $A^\kappa,\tilde A_\kappa$ & $\kappa = 1, \ldots ,h^{2,1}_+ (Y_3)$ \\
			\hspace*{1.2cm} $H^3_-(Y_3)$ & $(\alpha_{\hat k}, \beta^{\hat k})$ &  $z^k$ & $\hat k = 1, \ldots ,h^{2,1}_- (Y_3)+ 1 $ \\ 
			\bottomrule
		\end{tabular}
		\caption{Real basis for the cohomology groups. The dimensions 
			are denoted by $h^{p,q}_\pm (Y_3)= \text{dim}\, H^{p,q}_\pm(Y_3)$. 
			$(\alpha_\kappa, \beta^\kappa)$ and $(\alpha_{\hat k}, \beta^{\hat k})$ are symplectic basis.
			Our index conventions include $k=1,\ldots, h^{2,1}_-$, while the 
			hat on $\hat k$ indicates the labeling of one further element. We also list the four-dimensional 
			fields associated to these basis elements in the expansions \eqref{exp_formfields}.} \label{BasisForms}
	\end{center}
\end{table}
This leads to the following expansion 
of the K\"ahler form $J$ of $Y_3$, and the NS-NS and R-R form fields
\begin{equation}
\begin{aligned}
\label{exp_formfields}
J&=v^\alpha \omega_\alpha\, , \quad B_2 = b^a \omega_a\, , \quad  C_2 = c^a \omega_a\, , \\ 
C_4 &= C_2^\alpha \wedge \omega_\alpha + \rho_\alpha \tilde \omega^\alpha + A^\kappa \wedge \alpha_\kappa + \tilde A_\kappa \wedge \beta^\kappa \, ,  
\end{aligned}
\end{equation}
where $c^a$, $b^a$, and $\rho_\alpha$ are scalars, $C_2^\alpha$ are two-forms, and 
$(A^\kappa,\tilde A_\kappa)$ are vectors in the four-dimensional effective theory. 
It is crucial to stress that $C_4$ has a self-dual field-strength, given by $F_5 =\dd C_4 + \frac{1}{2} B_2 \wedge \dd C_2-\frac{1}{2} C_2 \wedge \dd B_2$.\footnote{Notice that we use a different convention than~\cite{Jockers:2004yj} for the field $C_4$. In particular, they are related by $C_4^{\text{here}}= C_4^{\text{there}}-\frac12 B_2\wedge C_2$. In order to compare with the results obtained from the F-theory reduction, it is more convenient to use this convention, which makes $C_4^{\text{here}}$ invariant under $Sl(2,\mathbb Z)$.}
This yields a duality 
between the two-forms $C_2^\alpha$ and scalars $\rho_\alpha$, and identifies
$\tilde A_\kappa$ as the magnetic dual of $A^\kappa$.
Therefore, we can eliminate the 
two-forms $C^\alpha_2$ in favor of $\rho_\alpha$ and 
the vector $\tilde A_\kappa$ in favor of $A^\kappa$. 
It is, however, interesting to point out that the structures 
we discuss later on can be also analyzed in the  
dual frames as we will see in section~\ref{Fthy}. 
In addition to the zero modes of the forms~\eqref{exp_formfields},
also the axio-dilaton $\tau = C_0 + i e^{-\phi}$ reduces to a four-dimensional field. 
Finally, the deformations of the Calabi-Yau metric compatible with $\sigma$
are the K\"ahler structure deformations $v^\alpha$ 
and the complex structure deformations $z^k$
parameterizing forms in $H^{2,1}_-(Y_3,\bbC)$. Note that $\tau$ and $z^k$ 
are complex fields. 

Before turning to the D7-branes let us note that a general $\cN=1$ compactification 
can include background fluxes $H_3$ and $F_3$~\cite{Grana:2005jc,Douglas:2006es}. These transform negatively 
under $\sigma^*$ and therefore admit an expansion
\beq
H_3 =  m_{H}^{\hat k} \alpha_{\hat k} + e^H_{\hat k} \beta^{\hat k} \, ,\qquad F_3 =  m_F^{\hat k} \alpha_{\hat k} + m^F_{\hat k} \beta^{\hat k} \, ,
\eeq 
with the basis introduced in table~\ref{BasisForms}.
It is well-known that these fluxes induce a non-trivial superpotential in this Type IIB setting~\cite{Gukov:1999ya}.
In the following we will not discuss background fluxes in much detail. While they can 
be included in the bulk sector without much effort, we will require however that 
they do not alter the couplings of the D7-brane.  

The coupling to a single 
space-time filling D7-brane was studied in detail in~\cite{Jockers:2004yj,Jockers:2005zy} by dimensionally reducing 
the D7-brane Born-Infeld and Chern-Simons actions. In order to review the 
results we will make some simplifying assumptions. In particular, we will analyze  
on the dynamics of a single D7-brane while being aware that a tadpole 
canceling configuration requires the inclusion of other D7-branes.\footnote{A more thorough discussion of 
	the global constraints on such settings can be found, for example, in~\cite{Blumenhagen:2008zz}.
	We refer the reader to these works especially for the discussion of the D5-brane tadpole 
	constraint and the appropriate quantization conditions.}
This will allow us to focus on the 
structures relevant to this work. Some interesting generalizations will appear in 
the study of the F-theory vacua of section~\ref{Fthy}. In particular, the F-theory analysis 
contains the proper inclusion of the seven-brane deformation (or position) moduli. 

Let us consider a D7-brane 
wrapped on a divisor $S$ in $Y_3$ and denote its orientifold image by $S'=\sigma(S)$. 
It is useful to introduce $S_+ = S \cup \sigma(S)$ and $S_-= S \cup -\sigma(S)$, where the minus sign 
stands for orientation reversal. This allows to split the cohomologies 
$H^{p,q}(S_+) =H^{p,q}_+(S_+) \oplus H^{p,q}_-(S_+)$ under $\sigma$. Then, the eight-dimensional gauge field $A$ and embedding $\zeta$ of the D7-brane image pair can be 
expanded as~\cite{Jockers:2004yj,Jockers:2005zy}
\begin{alignat}{3} \label{AD7_exp}
A &= A_{\rm D7}P_- + a_p \gamma^p + \bar a_p \bar \gamma^p\, , \qquad &p &= 1,\ldots,h_-^{1,0}(S_+)\,,\\
\zeta &=  \zeta_K \, s^K + \bar \zeta_K \, \bar s^K\, ,  \qquad\quad \qquad &K &= 1,\ldots,h_-^{2,0}(S_+)\,,
\end{alignat}
where $P_-$ is a function equal to $+1$ on $S$ and $-1$ on $\sigma(S)$. The fact that these have to be expanded
into $H^{1,0}_-(S_+)$ and $H^{2,0}_-(S_+)$, respectively, follows from the action of the orientifold on
the open string states.  

It is important to stress that the notion of $\gamma^p$ being $(0,1)$ implies that the forms depend on 
the complex structure moduli $z^k$ of the ambient Calabi-Yau space $Y_3$. 
To make this dependence more explicit, we can expand 
\beq \label{gammaI_def}
\gamma^p = \frac 12 \rf^{pq} (\hat \alpha_q - i \wb f_{qr} \hat \beta^r)\, ,
\eeq
where $(\hat \alpha_p,\hat \beta^p)$ is a real basis of $H^{1}(S)$. 
Here $f_{pq}$ is a holomorphic function in the complex structure moduli $z^k$. For an appropriate 
basis, its real part $\rf_{pq}$ is invertible and we denote the inverse by $\rf^{pq}$. 
This ansatz can be justified in the F-theory reduction as argued in 
\cite{Grimm:2010ks,Grimm:2015ona,Greiner:2015mdm} and 
was recently used in Type IIB orientifolds in~\cite{Carta:2016ynn}. While not a priori obvious 
a parametrization of the form \eqref{gammaI_def} will allow us to bring the effective action into standard $\cN=1$ 
form.  
This is most clearly seen in the F-theory treatment to which we will come back  
in section~\ref{Fthy}. Clearly, one can also expand $A$ into the real basis 
$(\hat \alpha_p,\hat \beta^p)$ such that 
\begin{equation}
\begin{aligned}
\label{AD7_exp2}
A &= A_{\rm D7} P_- + \tilde c^{\,p} \, \hat \alpha_p  +  c_p \, \hat \beta^p \, ,\\
a_p &= i c_p + f_{pq} \, \tilde c^{\,q}\, .
\end{aligned}
\end{equation}
The basis $(\hat \alpha_p,\hat \beta^p)$ is independent of the complex structure deformations and 
therefore all complex structure dependence in $a_p$ is again captured by the function $f_{pq}$.
We summarize our notation for the open string sector in table~\ref{openBasisForms}.

\begin{table}[h!]
	\begin{center}
	\begin{tabular}{cccl}
	\toprule 
	cohomology group  \qquad& basis element &  \qquad fields \qquad \qquad& index range  \\
	\midrule
	$H_-^{1,0}\, (S_+)$  & $\gamma^p$ & $a_p$ & $ p = 1, \ldots h^{1,0}_- \, (S_+)$\\
	$H_-^{2,0}\, (S_+)$ & $s^K$ & $\zeta_K$ & $ K = 1, \ldots h^{2,0}_- \, (S_+)$ \\
	$H_-^{1}\, (S_+)$  & $(\hat \alpha_p,\hat \beta^p)$ & $(\tilde c^p, c_p)$& $ p = 1, \ldots h^{1,0}_- \, (S_+)$ \\
	\bottomrule
	\end{tabular}
		\caption{Cohomology groups on the D7-brane divisor $S_+$. The dimensions 
			are denoted by $h^{p,q}_\pm = \text{dim}\, H^{p,q}_\pm$. While $\gamma^p$ and $s^K$ are 
			complex basis elements, the forms $(\hat \alpha_p, \hat \beta^p)$ constitute a real basis.} \label{openBasisForms}
	\end{center}
\end{table}

We are now in the position of stating our simplifying assumptions. First, we will assume that\footnote{Here and 
	in the following we will denote by $[D]$ the two-form class Poincar\'e dual to the divisor $D$.} 
\beq
[\sigma(S)] = [S]\, ,
\eeq
i.e.~that $S$ and its orientifold image $S'$ are in the same homology class. 
This implies that the $U(1)$ gauge field of the D7-brane is not massive by 
a geometric St\"uckelberg mechanism~\cite{Jockers:2004yj,Grimm:2011tb,Braun:2014nva}. 
And second, we will assume the vanishing of the intersections 
\beq \label{flux-vanish}
\int_{S_+} i^* \alpha_{\hat k} \wedge \hat \alpha_p =  \int_{S_+} i^* \alpha_{\hat k} \wedge \hat \beta^p = 
\int_{S_+} i^* \beta^{\hat k} \wedge \hat \alpha_p =  \int_{S_+} i^* \beta^{\hat k} \wedge \hat \beta^p = 0\, , 
\eeq
where $i$ denotes the embedding map of $S_+$ into $Y_3$, $i: S_+  \hookrightarrow Y_3$. This condition ensures 
that there is no superpotential that obstructs complex structure and Wilson line deformations.\footnote{This was discussed in~\cite{Jockers:2004yj} from the perspective of relative cohomology and was derived in~\cite{Marchesano:2014iea} from backreaction effects in supergravity.}
The considered D7-branes can admit an arbitrary number $h_-^{2,0}(S_+)$ of deformations $\zeta_K$ 
and $h_-^{1,0}(S_+)$ of Wilson line moduli $a_p$. To keep the presentation simple, we will freeze the fields $\zeta_K$ 
as well as all matter fields arising at the intersections among D7-branes. 
This will allow us to focus the following discussion the couplings of the Wilson line moduli $a_p$. In the F-theory reduction, presented in section~\ref{Fthy}, 
a general dependence on the seven-brane deformations will be included and also charged matter states are (implicitly) accounted for.

Let us note that the condition \eqref{flux-vanish} is only imposed for the orientifold negative forms $(\alpha_{\hat k},\beta^{\hat k})$
in $Y_3$. The positive forms $(\alpha_{\kappa},\beta^\kappa)$ can non-trivially intersect the negative one-forms on 
$S_-$. 
Thus, we introduce the intersection numbers
\begin{equation}
\begin{alignedat}{3} \label{Mkinetic}
M_{\kappa p} &= \int_{S_-} i^* \alpha_{\kappa} \wedge \hat \alpha_p \, , \qquad 
&M_{\kappa}{}^p & = \int_{S_-} i^* \alpha_{\kappa} \wedge \hat \beta^p \, ,\\
M_p{}^{\kappa} &= \int_{S_-} \hat \alpha_p\wedge i^* \beta^{\kappa} \, , \qquad 
& M^{p \kappa} &= \int_{S_-} \hat \beta^p\wedge  i^* \beta^{\kappa} \, .
\end{alignedat}
\end{equation}
As we discuss in subsection~\ref{TypeIIBmixing}, these couplings control the kinetic mixing 
of the D7-brane $U(1)$ $A_{\rm D7}$ with the R-R gauge fields $A^\kappa$ of the bulk theory.

We are now in the position to display the four-dimensional $\cN=1$ complex coordinates. 
First, we have the complex fields 
\beq \label{Set1}
\text{\underline{Set 1:}}\qquad   \tau=C_0 + i e^{-\phi}, \quad z^k\, ,
\eeq 
which are already complex in our reduction ansatz. Their complex 
structure does not depend on other fields in the reduction. Note that 
the D7-brane deformations $\zeta_K$ are part of Set 1, but have been frozen to 
keep the presentation simpler. 
Second, there are the complex fields 
\beq \label{Set2}
\text{\underline{Set 2:}} \qquad G^a = c^a - \tau b^a \, , \quad a_p\, ,  
\eeq
which admit a complex structure that changes with the values of the fields in 
Set 1 given in \eqref{Set1}. This is obvious from the definition of $G^a$ 
and readily inferred for the $a_p$'s by noting that they are coefficients of 
complex structure dependent $(0,1)$-forms in \eqref{AD7_exp}.
Finally, there is a third set of fields:
\beq \label{Set3}
\text{\underline{Set 3:}}\qquad T_\alpha   = \frac{1}{2}  \cK_{\alpha \beta \gamma} v^\beta v^\gamma + i \rho_\alpha + \frac{i}{2(\tau -\bar \tau)} \cK_{\alpha ab} G^a (G-\bar G)^b + \frac{1}{2} d_{\alpha}{}^{pq} a_p (a + \bar a)_q\, ,
\eeq
which non-trivially depends on the fields in Set 1 and Set 2.
The $T_\alpha$ are often termed the complexified K\"ahler structure moduli.
The introduced couplings are given by the $Y_3$ intersection numbers
\beq
\cK_{\alpha \beta \gamma} = \int_{Y_3} \o_\a \we \o_\b \we \o_\g \, , \qquad
\cK_{\alpha ab}  = \int_{Y_3} \o_\a \we \o_a \we \o_b \, , 
\eeq
as well as the complex structure dependent function
\beq
d_{\alpha}{}^{pq} = i \int_{S_+} i^* \omega_\alpha \wedge \gamma^p \wedge \bar \gamma^q 
= - \frac{1}{2} \rf^{qr} Q_{\a r}{}^p\,, \qquad  Q_{\a r}{}^p =  M_{\alpha r}{}^p + i f_{rs} M_{\alpha}{}^{sp}\,,
\eeq
with 
\begin{equation} 
\label{def-MQ}
M_{\alpha p}{}^q = \int_{S_+} i^* \omega_\alpha \wedge  \hat \alpha_p \wedge \hat \beta^q\, , \qquad 
M_{\alpha}{}^{pq} = \int_{S_+} i^* \omega_\alpha \wedge  \hat \beta^p \wedge \hat \beta^q\,.
\end{equation}

For completeness, let us note that the K\"ahler potential takes the seemingly simple form 
\beq \label{IIBKpot}
\hat K = - 2 \log \cV - \log (\tau - \bar \tau ) - \log \Big( \int_{Y_3} \Omega \wedge \bar \Omega\Big)\, . 
\eeq
This K\"ahler potential depends on the complex coordinates \eqref{Set1}-\eqref{Set2}, i.e.~we 
identify in \eqref{act4D} that 
\beq
   \hat M^\bA = (\tau,\,z^k,\,G^a,\, a_p,\, T_\alpha)\, . 
\eeq
All the field dependence of this $\hat K$ on the 
fields of Set 2, i.e.~the $G^a$ and $a_p$, arises only through the definition of $T_\alpha$. 
In fact, we note that the volume $\cV= \frac{1}{6} \cK_{\alpha \beta \gamma} v^\alpha v^\beta v^\gamma$ 
in \eqref{IIBKpot} depends on $T_\alpha$ by solving \eqref{Set3} for $v^\alpha$, which then introduces 
a dependence on $G^a,a_p$ mixed with $\tau,z^k$.

To conclude this subsection we discuss a special case for the above compactification 
separately in which several of the couplings simply. More precisely, we briefly 
summarize the above result for $h^{1,0}_-(S_+) = 1$ and $h^{1,1}_-(Y_3)=0$, i.e.~the case in which  
the rigid D7-branes only admits a single complex Wilson line modulus $a$. In this case the 
dynamics of $a$ is encoded by the correction to $T_\alpha$ given by 
\beq \label{Talpha_special}
T_\alpha   = \frac{1}{2}  \cK_{\alpha \beta \gamma} v^\beta v^\gamma + i \rho_\alpha  - \frac{1}{4} (\rf)^{-1} M_\alpha\, a (a + \bar a)\, ,
\eeq
where we have used that $M_{\alpha p}{}^q$ in \eqref{def-MQ} reduces to a vector denoted by $M_\alpha$ and that $M_{\alpha}{}^{pq}$
vanishes due to antisymmetry for one modulus. The kinetic terms of $a$ depend non-trivially  on the 
complex structure moduli $z^k$ through the holomorphic function $f$.

\subsection{Continuous and discrete shift-symmetries}

Having introduced the complex coordinates \eqref{Set1}, \eqref{Set2}, and \eqref{Set3}
we are now in the position to discuss the symmetries. 
In order to do that we first recall that $G^a$ and $T_\alpha$ 
contain zero modes of R-R and NS-NS forms and therefore  
inherit discrete symmetries from large gauge transformations of $C_2$, $B_2$ and $C_4$. These are shifts by \emph{integral} closed 2-forms, namely
\beq \label{deltaC2B2}
\delta C_2 = \lambda^a \omega_a \, , \qquad  \delta B_2 = \tilde \lambda^a \omega_a\, ,
\eeq
where $\lambda^a$ and $\tilde \lambda^a$ are appropriately quantized constants.\footnote{As usual, the four-dimensional theory
	obtained from dimensional reduction is invariant under a continuous version of the symmetry, while quantum
	effects break it to the discrete subgroup.} Turning to $C_4$, an obvious large gauge transformation is $\delta C_4 = \lambda_\alpha \tilde \omega^\alpha$, for 
constant $\lambda^\alpha$. However, we note that the field-strength 
$F_5 =\dd C_4 + \frac{1}{2} B_2 \wedge \dd C_2-\frac{1}{2} C_2 \wedge \dd B_2$ actually contains terms 
depending on $C_2$ and $B_2$.  Therefore, the shifts \eqref{deltaC2B2} induce a 
shift of $C_4$ as
\beq \label{deltaC4}
\delta C_4 =  \lam_\a \, \tilde \omega^\alpha - \frac{1}{2}\tilde \lam^a \omega_a \wedge C_2 + \frac{1}{2} \lam^a \omega_a \wedge B_2  \, .  
\eeq
A second set of symmetries arises from internal gauge transformations 
on the D7-brane world-volume. For constants $\lambda^p,\tilde \lambda_p$ these are 
parameterized by 
\beq \label{detlacA}
\delta A = \tilde \lambda^p \, \hat \alpha_p + \lambda_p \, \hat \beta^p\, .
\eeq
Also in this case one finds that the four-form $C_4$ has to shift. 
While we will not give the transformation of $C_4$ directly, 
let us point out that it can be inferred by noting the NS-NS two-form $B_2$ naturally 
combines with $F = \dd A$ on the D7-brane world-volume as 
\beq \label{def-cF}
\mathcal{F} = i^*B_2 - 2\pi \alpha' {F}\, ,
\eeq 
where we have temporarily restored the $\alpha'$ dependence. 
This implies that one can capture the gauge degrees of freedom of an Abelian D-brane with $B_2$, and the fact that the field $C_4$ shifts
under \eqref{detlacA} is already contained in \eqref{deltaC4}. 
A more detailed discussion how this is done in practice can be found in~\cite{Douglas:2014ywa}.
The transformations can be simply inferred when investigating the  
$\cN=1$ coordinates as we will see next. Furthermore, since $\mathcal F$ is gauge invariant,
under a shift of the B-field \eqref{deltaC2B2}, we have to shift the worldvolume flux on the brane
accordingly
\beq\label{deltaFD7}
\delta F=\frac{1}{2\pi \a'}\tilde \lambda^a i^*\omega_a\,.
\eeq

To examine the shifts of the $\cN=1$ chiral coordinates, we first focus on the fields of Set 2 defined in \eqref{Set2}. 
Performing the transformations \eqref{deltaC2B2} and \eqref{detlacA} we find that
\beq \label{deltaGa}
\delta G^a = \lam^a-\tau \tilde \lam^a\, , \qquad \delta a_p = i \lambda_p + f_{pq} \tilde \lambda^q\, , 
\eeq
where we have used that $a_p$ arises in the expansions \eqref{AD7_exp} and \eqref{AD7_exp2}. 
Both shifts are holomorphic 
in the moduli of Set 1 given in \eqref{Set1} and are shown to unify when using the F-theory description 
in terms of a Calabi-Yau fourfold (see section~\ref{Fthy}).
The fields of Set 3 have the most involved transformation properties:
\beq \label{deltaTalpha}
\delta T_\a =i \lam_\a- \frac{i}{2} \mathcal K_{\a ab}\tilde \lam^a (2G^b+\delta G^b) -\frac12 \tilde \lam^p Q_{\a p}{}^q(a_q+\delta a_q) -\frac12 a_q(M_\a{}^{pq}\lam_p+M_{\a p}{}^q\tilde \lam^p )\,,
\eeq
which can be inferred by investigating the isometries of the K\"ahler manifold spanned 
by all complex fields with K\"ahler potential \eqref{IIBKpot}. Notice that this is valid for finite values
of the transformation parameters and that the shift is holomorphic. 
It is also important to stress that  \eqref{deltaTalpha}
implies that the shift in $\delta \rho_\a$ not only depends on $\lam_\a$ but also on 
$\lambda^a$, $\tilde \lambda^a$, $\lambda_p$, $\tilde \lambda^p$.
As mentioned above, this is a consequence of the transformation rule for $C_4$ given in \eqref{deltaC4}, together with the shift induced by \eqref{detlacA}.
This, in turn, implies that the isometry group generated by the transformation is 
actually a non-Abelian. To see this, we introduce the Killing vectors $t_a$, $\tilde t_a$, $t^p$, $\tilde t_p$ and $t^\alpha$ 
for the symmetries parameterized by $\lambda^a$, $\tilde \lambda^a$, $\lambda_p$, $\tilde \lambda^p$, and $\lambda_\alpha$.
These are then found to respect the non-trivial commutators~\cite{Grimm:2015ona}
\beq \label{Heisenberg}
[t_{a}, \tilde t_{b}] = - \cK_{\alpha a b} \, t^\alpha \, , \qquad [t_p, \tilde t^q]= - M_{\alpha p}{}^{q} \, t^\alpha\, . 
\eeq
This algebra is a generalization of the well-known Heisenberg algebra. It is an interesting 
challenge to gauge this algebra while preserving supersymmetry~\cite{Hull:1985pq,Grimm:2015ona}.

As mentioned earlier, in the absence of gaugings for the isometries \eqref{deltaGa} and \eqref{deltaTalpha},
one expects that the continuous global shift-symmetries are actually broken to 
discrete symmetries at the quantum level. Since the discrete version of the symmetries comes from large gauge transformations in the
higher-dimensional $p$-form fields, such shifts actually identify field configurations
in the Set 2 to parameterize complex tori $\mathbb{T}^{2h^{1,1}_-}_{\rm closed}$ and $\mathbb{T}^{2h^{1,0}}_{\rm open}$, e.g.~one finds the identifications
\begin{equation}
\begin{alignedat}{3}
\label{cb-periods}
c^a&\simeq c^a+1\, , \qquad &b^a&\simeq  b^a+1\, ,\\
c_p&\simeq c_p+1\, , \qquad &\tilde c^p&\simeq  \tilde c^p+1\, ,
\end{alignedat}
\end{equation}
and $G^a$, $a_p$ parameterizing\,\footnote{We are sloppy here by assuming that the $\sigma^*$ split is compatible with restricting 
	to integer homology and by neglecting cohomological torsion.} 
\beq \label{complex-tori}
\mathbb{T}^{2h^{1,1}_-}_{\rm closed}  = \frac{H^{1,1}_-(Y_3,\bbC)}{H^{2}_-(Y_3,\bbZ)}\, , \qquad  \mathbb{T}^{2h^{1,0}_-}_{\rm open}  = \frac{H^{1,0}(S,\bbC)}{H^{1}(S,\bbZ)}\,.
\eeq 
The complex structure on $\mathbb{T}^{2h^{1,1}_-}_{\rm closed} $ is simply given by $\tau$, while the complex structure on $\mathbb{T}^{2h^{1,0}}_{\rm open}$ is encoded in the holomorphic function $f_{pq}$.
Finally, also $\rho_\alpha$ is periodic $\rho_\a\simeq  \rho_\a+1$, but one has 
to additionally impose identifications under \eqref{cb-periods} using $\delta \rho_\alpha$ obtained from \eqref{deltaTalpha}.
These identifications render the field space spanned by $c^a,b^a,c_p,\tilde c^p$ and $\rho_\alpha$ to be 
compact.

\subsection{The $\cN=1$ gauge coupling function}\label{sec:IIBcoupling}

We turn now to the $\cN=1$ gauge coupling function for the 
Type IIB orientifold setting and study its symmetries. To keep the 
discussion simple, we first focus on the case in which the kinetic mixing is absent, 
i.e.~the case in which the couplings \eqref{Mkinetic} are zero
\beq\label{no_kin}
M_{\kappa p} =  M_{\kappa}{}^p =M_p{}^{\kappa}=M^{p\kappa }=0\, .
\eeq 
We will comment on the more general situation in subsection~\ref{TypeIIBmixing}.

A first way to obtain the gauge coupling function is by performing a direct 
dimensional reduction. For the R-R gauge fields $A^\kappa$ one then 
finds~\cite{Grimm:2004uq}
\beq \label{RR-gaugecoupling}
\hat f_{\kappa \lambda} = \cF_{\kappa \lambda} |_{z^\kappa=0}\, ,
\eeq
where $\cF_{\kappa \lambda} = \partial_{z^\kappa}\partial_{z^\lambda} \cF$ is the 
second derivative of the holomorphic $\cN=2$ pre-potential $\cF$ for $Y_3$ of the underlying 
theory. The restriction in \eqref{RR-gaugecoupling} is to the slice 
of complex structure deformations that are compatible with the orientifold 
condition $\sigma^* \Omega = - \Omega$, for the $(3,0)$-form of $Y_3$.
The function $\hat f_{\kappa \lambda} $ is  thus holomorphic in 
the complex structure deformations $z^k$.  

Let us next include the D7-brane. In the absence of the moduli $G^a$ and  $a_p$ in Set 2, one 
finds by a reduction of the Dirac-Born-Infeld and Chern-Simons action that
\beq\label{noflux}
\hat f_{\rm D7} = \delta^\alpha_{\rm D7} \Big(\frac{1}{2} \cK_{\alpha \beta \gamma} v^\beta v^\gamma + i \rho_\alpha \Big)\, . 
\eeq
Here $\delta^\alpha_{\rm D7}$ is the restriction to the world-volume $S_+$ 
and can be obtained by expanding the Poincar\'e-dual two-from $[S_+]$ to $S_+$ into 
the basis $\omega_\alpha$, i.e.~$[S_+] = \delta^\alpha_{\rm D7} \, \omega_\alpha$.
The real part of $\hat f_{\rm D7}$ is determined by using the calibration conditions 
for supersymmetric cycles and thus obtained from the volume of $S_+$ measured in the
ten-dimensional Einstein-frame metric. In the string frame one has $\Re  \hat f_{D7} \propto g_s^{-1}$.
Clearly, in the absence of fields of Set 2 the gauge-coupling is $\hat f_{\rm D7} =\delta^\alpha_{\rm D7} \, T_\alpha$
and thus holomorphic in the $\cN=1$ coordinates. Its imaginary part 
non-trivially shifts with $\lambda_\alpha$ under \eqref{deltaC4}, which are the standard constant shifts of the 
theta-angle. 

The inclusion on the $G^a$ moduli is also straightforward, since the corrections in $G^a$ are 
at the same order of $g_s$ as the volume part. Indeed, dimensionally reducing the D7 action
one finds that, with vanishing worldvolume flux, the gauge coupling function is~\cite{Jockers:2004yj}\footnote{The slightly odd factor of $1/2$ in the term proportional to $\cK_{\alpha ab}c^a b^b$
	arises due to fact that our $C_4$ is shifted such that it is $Sl(2,\bbZ)$ invariant in Type IIB.}
\beq\label{fD7reduction}
\hat f_{\rm D7} = \delta^\alpha_{\rm D7} \left [\frac{1}{2} \cK_{\alpha \beta \gamma} v^\alpha v^\beta 
+ \frac{1}{2}  e^{-\phi}\cK_{\alpha ab}   b^a b^b  + i \left (\rho_\alpha - \frac{1}{2} \cK_{\alpha ab} c^a b^b + 
C_0 \frac{1}{2} \cK_{\alpha ab} b^a b^b \right) \right]\, ,
\eeq
which is holomorphic in the $T_\alpha$ coordinates \eqref{Set3} in the absence of Wilson line moduli. 
We note that, naively, the gauge coupling function is now transforming non-trivially under 
the symmetries \eqref{deltaTalpha} since, in addition to the constant shifts with $\lambda_\alpha$,
one also finds shifts with $\lambda^a$ holomorphic in $G^a$ and $\tau$. However, \eqref{fD7reduction}
is only valid when the gauge flux on the D7-brane is zero, i.e.~$F=0$, which as noted above, is not a gauge invariant
condition since it shifts according to eq.~\eqref{deltaFD7}. Thus, the gauge invariant version of \eqref{fD7reduction} is actually
\beq\label{fD7flux}
\hat f_{\rm D7} = \delta^\alpha_{\rm D7}\left ( T_\a +i\cK_{\a a b}f^a G^b+\frac{i}{2}\tau \cK_{\a ab}f^a f^b   \right )\,,
\eeq
where we defined the worldvolume fluxes $f^a$ as
\beq
F=\frac{1}{2\pi \a'}f^ai^*\omega_a.
\eeq
Since these transform according to \eqref{deltaFD7}, we find that the gauge coupling function is both holomorphic and invariant under the
whole set of shift symmetries (modulo a constant imaginary shift), as it should.

Finally, when including the Wilson line moduli for the D7-brane, we immediately face a problem. At first, one might think that the
gauge coupling function is given in this case by \eqref{fD7flux}, where $T_\a$ contains a quadratic term in the Wilson lines \eqref{Set3}.
However, the dimensional reduction of the D7-brane action does not give such a term and we find again \eqref{noflux}. As argued in~\cite{Jockers:2004yj}, a contribution
quadratic in the Wilson lines is generated at one loop in $g_s$ and is therefore natural that it is not captured by the Dirac-Born-Infeld action,
which is only valid at tree level in open string amplitudes.\footnote{This was also noticed in the mirror dual configurations~\cite{Grimm:2011dx},
which were also studied in \cite{Kerstan:2011dy}. Corrections in Type IIA orbifolds have been studied, for example, in \cite{Lust:2003ky,Gmeiner:2009fb,Honecker:2011sm}.} 
Such corrections were computed in~\cite{Berg:2004ek,Berg:2004sj} in toroidal models,
which show that indeed, a quadratic term arises at one loop level. 
It is therefore natural to split $\hat f_{\rm D7}$ as
\beq \label{fsplit_class_oneloop}
\hat f_{\rm D7}= \hat f_{\rm D7}^{\rm red} + \hat f_{\rm D7}^{\rm 1-loop} \, ,
\eeq
where $\hat f_{\rm D7}^{\rm red}$ is obtained by direct dimensional reduction of the D7-brane action. 
Comparing \eqref{fsplit_class_oneloop} with \eqref{fD7flux} one is lead to make the ansatz
\beq \label{fone-loop}
\hat f_{\rm D7}^{\rm 1-loop} =  \frac{1}{2} \delta^\alpha_{\rm D7} \, d_{\alpha}{}^{pq} a_p (a_q + \bar a_q) + \log \Theta \, ,
\eeq
where $\Theta$ is a holomorphic function.
Note that our analysis of the shift symmetries implies that the quadratic term in \eqref{fone-loop} 
cannot be the full result, since under shifts of the Wilson line moduli, the field $T_\a$ shifts by a non-constant term, which would make the gauge coupling function
non-gauge invariant. We therefore introduced the non-vanishing holomorphic function 
$\Theta$ in the moduli $a_p$ and $z^k$. In the next section we discuss the properties of this completion in more detail.

\subsection{One-loop corrections and theta-functions} \label{sec:theta-ori}

Let us have a closer look at the inclusion of the Wilson line moduli in the 
discussion of the D7-brane gauge coupling function $\hat f_{\rm D7}$. 
As stressed above the quadratic term in the $a_p$ arise at order 
$g_s^0$, i.e.~is only visible at the open string one loop level. 
In toroidal models~\cite{Berg:2004ek,Berg:2004sj} it was furthermore shown 
that the fully corrected gauge coupling function contains a 
Riemann theta function depending on the D-brane moduli. In 
toroidal models, these theta functions arise due to the underlying 
toroidal compactification space. While we are not dealing 
with such a simple geometry, we have stressed in \eqref{complex-tori}
that the Wilson lines in this more general orientifold compactification 
also parameterize a higher-dimensional complex torus. 
In the following we will use this fact 
together with the transformation property \eqref{deltaTalpha}
to infer the general form of $\hat f_{\rm D7}$ as a 
function of $a_p$. More precisely, we suggest that 
$ \Psi = e^{\hat f^{\rm 1-loop}_{\rm D7}}$ introduced in \eqref{fone-loop} can be viewed as a holomorphic 
section of a certain line bundle on the torus $ \mathbb{T}^{2h^{1,0}_-}_{\rm open}$ introduced 
in \eqref{complex-tori}. Our construction is inspired by the discussion 
of the M5-brane action first given in~\cite{Witten:1996hc}. It has been extended and applied
relevantly for our orientifold setting, for example, in ref.~\cite{Belov:2006jd,Belov:2006xj}. A similar strategy has been also suggested in the 
construction of the non-perturbative $\cN=1$ superpotential~\cite{Ganor:1996pe,Grimm:2007xm,Grimm:2011dj,Kerstan:2012cy}.

\subsubsection{A simple case with one Wilson line modulus}

Before discussing the general case let us exemplify our reasoning 
for a single  Wilson line $a$, i.e.~for the situation discussed around \eqref{Talpha_special}.
The complex field $a$ parameterizes a complex two-torus $\mathbb{T}^{2}_{\rm open}$
with complex structure given by the function $f$. As above we can write $a = i c + f \tilde c$ 
with $c \cong c + 1$, $\tilde c \cong \tilde c+1$.
We then introduce the following connection on this torus
\be\label{con}
\mathfrak{A}=\frac{iM}{4 \re f}( a \dd\bar a-\bar a \dd a)\, .
\ee
with $M\in2\pi \mathbb Z$. The field strength $\mathfrak{F}=\frac{i M}{2\re f}\dd a\wedge \dd\bar a$ is a (1,1)-form, so $\mathfrak{A}$ is a connection on a holomorphic line bundle $\cL$. Holomorphic sections of $\cL$ are defined as sections that satisfy
\be\label{holom_section}
\bar \p_{\mathfrak{A}}\,\Psi=\left (\bar \p-i  \mathfrak{A}_{\bar a}\right )\Psi=0\,,
\ee
where the differential is with respect to $\bar a$. 
Note that $\Psi$ is defined on a torus and thus has to respect appropriate 
boundary conditions. Compatibility 
of \eqref{holom_section} with the torus shifts $a \cong a + n i  +  m f $, with $n,m\in \bbZ$, implies that $\Psi$ has to transform as
\be\label{bound}
\Psi(a +  n i  +  m f )=\text{exp}\left ( -\frac{i M}{2\re f}\im\left [ \left (in + fm \right )\bar a\right ] \right)\Psi(a)\, ,
\ee
where we kept $f$ constant, therefore ignoring the dependence on complex structure.
One can now simply solve the differential equation \eqref{holom_section} together with the boundary conditions \eqref{bound}. There are $|M|/2\pi$ linearly independent solutions given by (see e.g.~\cite{Cremades:2004wa})\footnote{One can show that there are $|M|$ independent solutions without having to solve the equation.
	This follows from an index theorem, see e.g.~\cite{GSW}, which in this case is $\int_{T^2}\mathfrak F=-M$.} 
\be \label{Psij_solutions}
\Psi_j=e^{-\frac{ M}{4\text{Re} f}a(a+\bar a)}\vartheta\left [\begin{array}{c} \frac{2\pi j}{M} \\ 0\end{array}\right ]\left (\frac{iMf}{2\pi},iMa\right )\,,\qquad \quad j=0,1,\dots,|M|/2\pi-1 \,,
\ee
with
\be\label{deftheta}
\vartheta\left [\begin{array}{c} \mu \\ \nu\end{array}\right ](\tau,a) = \sum_{l\in \mathbb Z} e^{i\pi \tau (\mu+l)^2}e^{2\pi i (\mu+l)(a+\nu)} 
\ee
the Jacobi theta function. Notice that the theta functions above can be seen as holomorphic sections of the bundle defined by \eqref{con} in \emph{holomorphic gauge}, i.e. with $\mathfrak{A}^{0,1}=0$ but $\mathfrak{A}^{1,0}\neq 0$, defined by the following complex gauge transformation\footnote{Usually we consider only gauge transformations $A\rightarrow A+d\chi$ with $\chi$ a real function. However, the eq.~\eqref{holom_section} is invariant under the complexified gauge group so we may take $\chi$ complex.}
\be
\mathfrak{A}_{\rm h}=\mathfrak{A}-\dd\left ( \frac{i M}{2\re f}a\re a\right )=-\frac{ i M}{\re f}\re a\dd a\,.
\ee
One thus recovers the standard transformation behavior of the theta functions under the torus shifts. 

In order to relate the $\Psi_j$ given in \eqref{Psij_solutions} to the gauge coupling function we next consider taking the logarithm of 
an arbitrary solution $\Psi=\sum_{j=0}^{|M|-1}C_j \Psi_j$,
\be\label{log}
\log\Psi= -\frac{ M}{4\text{Re} f} \, a \, (a+\bar a)+\log \Theta\, , \qquad \Theta=\sum_{j=0}^{|M|-1} C_j\ \vartheta\left [\begin{array}{c} \frac{2\pi j}{M} \\ 0\end{array}\right ]\left (\frac{iMf}{2\pi},iMa\right )\, .
\ee
This equation is already quite illuminating. The first piece is precisely the correction to the $T_\a$ coordinate proportional 
to the moduli $a$, as in eq.~\eqref{Talpha_special}. The second term, $\log\Theta$, is holomorphic in $a$ and transforms precisely in the right way to render 
$\delta^{\alpha}_{\rm D7} \, T_\alpha+\log\Theta$ invariant under shifts in $a$. Therefore, identifying 
\beq
\hat f_{\rm D7} = \delta^{\alpha}_{\rm D7} \, T_\alpha + \log \Theta \, ,
\eeq
with $M = \delta_{\rm D7}^\alpha M_\alpha$ and appropriate $C_j$, yields a suitable completion of the gauge coupling function 
of a D7-brane. As promised, we have identified $ \Psi = e^{\hat f^{\rm 1-loop}_{\rm D7}}$ as a holomorphic section 
of a line bundle on a two-torus, when viewing the one-loop part of the $T_\alpha$ coordinates as functions of $a,\bar a$.

Note that we have only focused on the $a$-dependence of $\hat f_{\rm D7}$ in the above discussion. We know, however, that 
supersymmetry implies that $\hat f_{\rm D7}$ also has to be holomorphic in the complex structure moduli $z^k$. 
Indeed, we find that our construction appropriately yields such a holomorphic dependence through the theta functions 
$\vartheta$ in \eqref{log} due to the holomorphic function $f(z^k)$. In general, however, the coefficients $C_j$ 
can also depend holomorphically on the moduli $z^k$. This dependence is not constrained by our considerations
of shift-symmetries. It can be constrained by including further symmetries, such as monodromy symmetries in the complex structure moduli space, but considerations of this type are beyond the scope of this work. 

\subsubsection{The general case with several Wilson line moduli}

Let us now repeat the same arguments for the more general situation with several 
Wilson line moduli $a_p$. The first step consists of constructing the line bundle $\cL$ 
on $\mathbb{T}^{2h^{1,0}_-}_{\rm open}$, by defining an 
appropriate  connection. We 
do this by analyzing the general transformations \eqref{deltaTalpha} of $T_\a$ under 
the torus shifts. 
Then we can follow the same strategy as above to constrain the expected one-loop correction. 

We would like to consider a holomorphic function $ \Theta(z^k,a_p)$ such that under 
the shift \eqref{deltaGa} of the $a_p$ satisfies
\be
\Theta(z^k,a_p+\delta a_p)=  \text{exp}\big(-\delta^\a_{\rm D7}\, \delta T_\a \big)\ \Theta(z^k,a_p)\,,
\ee
with $\delta T_\a$ given in \eqref{deltaTalpha}. The existence of such a $\Theta$ implies 
that 
\beq
\hat f_{\rm D7} = \delta^\a_{\rm D7} \, T_\a + \log \Theta \, ,
\eeq
remains invariant. As above, when viewing $T_\alpha$ as functions of $a_p$ we can identify 
$ \Psi = e^{\hat f^{\rm 1-loop}_{\rm D7}}$ as a holomorphic section of the line bundle $\cL$ satisfying \eqref{holom_section}
for some connection $\mathfrak{A}$.  

It is easier to determine the connection $\mathfrak{A}$ in holomorphic gauge
which reads
\beq \label{conno}
\mathfrak{A}_{\rm h}=\frac{i}{4} \delta_{\rm D7}^\a \left (2  M_{\a p}{}^q \re f^{pr} \re a_r+ M_{\a}{}^{pq} a_p\right ) \dd a_q\,.
\eeq
Indeed, one checks that (freezing complex structure) the connection transforms as 
\be
\mathfrak{A}_{\rm h}(a_p +\delta a_p)=\mathfrak{A}_{\rm h} (a_p)+ \dd \chi \, ,\qquad\quad \chi=-i \delta^\a_{\rm D7}\, \delta T_\a\,.
\ee
The field strength of $\mathfrak{A}_{\rm h}$ is
\be
\mathfrak{F} =-\frac{i}{4} \delta_{\rm D7}^\a \, M_{\a p}{}^r\re f^{pq}\, \dd a_r\wedge \dd \bar a_q\,,
\ee
where we imposed that
\be \label{constrainingf}
\delta_{\rm D7}^\a \,  \Big( M_{\a}{}^{pq}+\re f^{r[p}M_{\a r}{}^{q]} \Big)=0
\ee
such that $F^{2,0}=F^{0,2}=0$. Notice that the field strength does not depend on $M_{\a}{}^{pq}$ which, in particular, means that the number of solutions of \eqref{holom_section} is independent of $M_{\a}{}^{pq}$.\footnote{This is related to the fact that in \eqref{Heisenberg} 
	the couplings $M_{\a p}{}^q$ (and not $M_{\a}{}^{pq}$) determine the structure constants of the isometries of the scalar manifold.} 
Note that the constraint \eqref{constrainingf} can actually always be satisfied for a single D7-brane when choosing 
a basis $(\hat \alpha_p, \hat \beta^p)$ in \eqref{gammaI_def} that is symplectic with respect to the inner product 
$\langle \alpha,\beta \rangle = \int_{S_+} \delta^\alpha_{\rm D7} \, \omega_{\alpha} \wedge \alpha \wedge \beta$.\footnote{Note that this inner 
	product can be degenerate on the full set $(\hat \alpha_p, \hat \beta^p)$.}

We can thus infer the form of the solution $\Theta$ is a sum over the Riemann theta functions
\beq \label{gen-theta}
\vartheta \left [\begin{array}{c} \mu^p \\ \nu_p\end{array}\right ](f_{pq},a_p) 
= \sum_{l^p\in  \Gamma} e^{i\pi f_{pq} (\mu^p+l^p)(\mu^q+l^q)}e^{2\pi i (\mu^p+l^p)(a_p+\nu_p)}\,,
\eeq
where $\Gamma$ is a $h^{1,0}_-$-dimensional integer lattice. As in the simpler case considered before,
the coefficients in this sum can be complex structure dependent and are not constrained
by the torus shift-symmetries. In fact, in this section we worked in a fixed complex structure of the Calabi-Yau
threefold. For a proper treatment of the dependence on complex structure moduli, we should consider a line bundle
over $\mathbb{T}^{2h^{1,0}_-}_{\rm open}$, which is itself fibered over the space of complex structures.

This concludes our discussion on the interplay between holomorphicity of the gauge coupling function and
its behavior under the shift-symmetries of the axions when there is no kinetic mixing among the open and closed
string gauge bosons.

\subsection{Comments on kinetic mixing and gaugings} \label{TypeIIBmixing}

Up to now we have assumed that the kinetic mixing between the open and closed string $U(1)$'s vanishes, c.f.~\eqref{no_kin}. In 
this section we comment briefly on how the presence of mixing changes the situation (see~\cite{Abel:2008ai,Marchesano:2014bia} for a discussion on
kinetic mixing in D-brane models from a different perspective). 

As shown in~\cite{Jockers:2004yj}, the mixing is controlled by the couplings defined in \eqref{Mkinetic}. In our notation, the result that one
obtains from reducing the D7-brane action is
\beq
\hat f_{\k \text{D7}} = \re f^{pq}\re f_{\k\l} (M_{q}{}^\l-i\bar f_{qr}M^{r\l})\, a_p\,.
\eeq
Since both $f_{\k\l}$ and $f_{pq}$ depend holomorphically on complex structure, we find that $\hat f_{\k \text{D7}}$ has a complicated dependence on the complex
structure moduli, which does not seem holomorphic. However, the M-theory computation done in the next section shows that there is an identity which proves that this quantity is actually holomorphic. Indeed, one can show that
\beq\label{identitymix}
\re f^{pq}\re f_{\k\l} (M_{q}{}^\l-i\bar f_{qr}M^{r\l})= M_\k{}^p+if_{\k\l}M^{\l p}\,,
\eeq
and so the mixing becomes
\beq\label{kinmix}
\hat f_{\k \text{D7}} =  \left (M_{\k}{}^p+if_{\k\l}M^{\l p}\right ) a_p\,,
\eeq
which is now manifestly holomorphic. Notice that from the type IIB perspective this is a highly non-trivial identity among (2,1)-forms in the internal space
and (0,1)-forms on the worldvolume of the brane. However, in the F-theory description, both of them lift to three-forms in the Calabi-Yau fourfold, where the identity
\eqref{identitymix} becomes obvious (see the discussion around \eqref{identitiesD}).

Now we can analyze how the kinetic mixing behaves under the shift symmetries of the axions $a_p$. Clearly, $\hat f_{\k \text{D7}} $ is not invariant, which might be a reason to 
think that this cannot be correct, or at least not the full result. However, the presence of mixing has an interesting consequence for the symmetries, which
implies that the gauge coupling function must transform non-trivially under shifts in the Wilson lines. Again, since this is most easily seen from the M/F-theory
description done in the next section, we will just quote the result here. Under a shift
\eqref{deltaGa}, we have that\,\footnote{Notice that the couplings that appear in \eqref{transAIIB} are not exactly those in \eqref{kinmix}. However, using the identity \eqref{identitymix} we see that the transformation of the vectors is trivial if and only if the kinetic mixing is zero. This was already observed in~\cite{Marchesano:2014bia}.}
\begin{align}\label{transAIIB}
\begin{split}
\delta A^\k &=\dd\Lam^\k -(M_{p}{}^{\k} \tilde \lambda^p +M^{p\k}\lambda_p)A_{D7} \,,  \\
\delta A_{D7} & = \dd\Lam_{D7}\,,
\end{split}
\end{align}
where we included the corresponding gauge transformations of the vectors, $\Lam^\k$ and $\Lam_{D7}$.
Thus, a shift in the axions induces a \emph{constant} change of basis in the space of $U(1)$'s, which mixes the open and closed gauge bosons.
For integer values of $\tilde\lam^p$ and $\lam_p$, we find that the change of basis for the vectors is also integral, as expected from charge quantization. This, in turn,
implies that the gauge coupling function has to depend on the Wilson lines and should not be invariant under the symmetries, unlike in the case where the mixing vanishes. We leave a more detailed discussion to subsection~\ref{sec:dualsymmetries}.

Let us close this section with some remarks about the interplay between the transformation \eqref{transAIIB} and the gauging of the isometries \eqref{Heisenberg} of the scalar manifold from a purely field-theoretical perspective. As we stressed earlier, the isometries of the scalar manifold are non-Abelian, while the gauge symmetry of the vectors is Abelian. This suggests that one cannot gauge such isometries without introducing extra vectors or structure. However, this is not the case, precisely because the vectors transform as in
\eqref{transAIIB}. Indeed, suppose that we gauge the isometries
\beq
X_A=\Theta_A{}^p \tilde t_p+\Theta_{Ap} t^p+\Theta _{A\a}t^\a\,,
\eeq
where $A$ runs over $\k$ and the D7-brane gauge boson, and $\Theta$ is the embedding tensor. This means that, under a gauge transformation, we have to perform a
shift in the corresponding axions, namely
\beq\label{link}
\tilde \lam^p=\Theta_{A}{}^p\Lam^A\,,\qquad \quad  \lam_p=\Theta_{A p}\,\Lam^A \,, \qquad \quad \lam_\a=\Theta_{A\a}\,\Lam^A\,.
\eeq
Thus, the parameters $\tilde \lam^p$ and $\lam _p$ are generically no longer constant and the transformation \eqref{transAIIB} is not simply a constant change of basis. Instead, using \eqref{link} it becomes
\begin{align}\label{transAIIBss}
\begin{split}
\delta A^\k &=\dd\Lam^\k -(M_{p}{}^{\k}\Theta_A{}^p +M^{p\k}\Theta_{A p}) \, A_{D7} \, \Lambda^A  \,, \\
\delta A_{D7} & = \dd\Lam_{D7}\,,
\end{split}
\end{align}
which can be readily recognized as the gauge transformation of a \emph{non-Abelian} gauge group. Thus, we see that the transformation \eqref{transAIIB} allows to gauge certain non-Abelian isometry starting with an Abelian gauge group.  Finally, since the resulting gauge group is non-compact and non-semisimple, the gauge coupling function cannot be constant~\cite{Hull:1985pq}, which fits nicely with what we find from the reduction.\footnote{Actually, as shown in~\cite{Grimm:2015ona}, the gauge coupling function does not depend on the gaugings.} See~\cite{BerasaluceGonzalez:2012vb,Grimm:2015ona,Grimm:2015ska} for more details on the gauging of such isometries.



\section{M-theory on Calabi-Yau fourfolds and the F-theory frame}
\label{Fthy}

In this section, we perform the dimensional reduction of M-theory on a smooth Calabi-Yau fourfold $Y_4$ without fluxes. Then, by restricting to the case in which $Y_4$ is elliptically fibered, we perform the necessary dualization to compare the resulting three-dimensional theory to the circle reduction of an arbitrary four-dimensional $\mathcal N=1$ supergravity theory. Let us note that this approach has 
been already successfully applied in previous works, see e.g.~\cite{Denef:2008wq,Grimm:2010ks,Grimm:2011tb,Grimm:2015ona}.
However, it is crucial to stress that the reduction and comparison that we present is the most general analysis carried out so 
far.\footnote{Also in comparison to~\cite{Greiner:2015mdm} we will drop simplifying assumptions.} In particular, we will cover the cases 
that capture kinetic mixing between R-R bulk and 7-brane gauge fields. 

\subsection{Dimensional reduction of M-theory on a smooth fourfold} \label{smoothreduction}

We begin our analysis by performing the dimensional reduction of eleven-dimensional supergravity 
on $Y_4$. Such reductions were performed already in~\cite{Haack:1999zv,Haack:2001jz,Berg:2002es}, and we will deviate from these 
works only by considering a more explicit ansatz for the $(2,1)$-forms on $Y_4$. 

The starting point is the bosonic part of eleven-dimensional supergravity given by  
\begin{equation}
S^{(11)}  = \frac{1}{2} \int \Big( 
\hat  R\,  \hat \star 1 - \frac{1}{2} \hat G \w \hat \star\, \hat G - \frac{1}{6} \hat C \w \hat G \w \hat G \Big) \, ,
\label{11dAction}
\end{equation}
where $\hat R$ is the eleven-dimensional Ricci scalar and $\hat G = \dd \hat C$ is the four-form 
field strength for the three-form $\hat C$. 
We will consider backgrounds of the form
\begin{align} \label{backr}
\begin{split}
\langle \dd \hat s^2\rangle& =  \eta_{\mu\nu} \dd x^\mu \dd x^\nu + 2g_{m\bar n}\dd y^m \dd y^{\bar n}\, , \\
\langle\dd \hat C\rangle&=0\, ,
\end{split}
\end{align}
where $g_{m\bar n}$ is a Calabi-Yau metric on the fourfold $Y_4$.
This choice of background ensures that the resulting effective theory is a three-dimensional $\mathcal N=2$ supergravity. 

The effective theory of interest include all massless fluctuations around the background solution \eqref{backr}.
The massless modes arising from fluctuations of the metric can be encoded in terms of the K\"ahler form $J$ expanded as
\begin{equation}\label{JandO}
J= v^\S \omega_\S\, ,\qquad \quad \S=1,\dots , h^{1,1}(Y_4)\, ,  
\end{equation}
where $\omega_\S$  form a basis of harmonic two-forms. 
The fields $v^\S$ are three-dimensional real scalar fields that parametrize the K\"ahler structure deformations of $Y_4$. 
We also have $h^{3,1}(Y_4)$ complex fields $z^\cK$, $\cK=1,\ldots,h^{3,1}(Y_4)$, that encode the complex structure deformations of $Y_4$.

The massless modes that come from fluctuations of the M-theory three-form $\hat C$ are given by
\begin{equation}
\hat C =  A^\S \wedge\omega_\S + N_\cA \Psi^\cA +  \bar N_\cA \bar\Psi^\cA   \, , \qquad \quad \cA=1,\dots, h^{2,1}(Y_4)\, ,
\label{3dAnsatz}
\end{equation}
where we introduced $\Psi^\cA$, a basis of harmonic $(1,2)$-forms. We note that $A^\S$ are 
three-dimensional vector fields and $N^\cA$ are three-dimensional complex scalars. Following~\cite{Grimm:2010ks}, we choose the following parametrization of the $(1,2)$-forms
\begin{equation} \label{Psi_ansatz}
\Psi^\cA = \frac12 \rf^{\cA\cB} (\alpha_\cB - i \wb f_{\cB\cC} \beta^\cC)\,,
\end{equation}
where $(\alpha_\cA,\,\beta^\cB)$ are a basis of integral harmonic real three-forms and $f_{\cA\cB}$ is holomorphic in complex structure. 
We also defined $\rf^{\cA\cB}$, which is the inverse of $\rf_{\cA\cB}$. Thus, $\hat G$ is given by 
\begin{equation} \label{Ghat}
\hat G  = \dd A^\S \w \omega_\S + DN_\cA\w \Psi^\cA + D\bar N_\cA\w \bar\Psi^\cA  \,,
\end{equation} 
with
\begin{equation}
DN_\cA = \dd N_\cA -  \re N_\cB \re f^{\cB\cC}\partial_\cK  f_{\cC\cA} \dd z^\cK\,,\qquad D\wb N_\cA =\wb{DN_\cA }\,.
\end{equation}
Note that we could have also chosen to use the real basis $(\alpha_\cA,\beta^\cB)$ in the expansion of 
$\hat C$. This would introduce real scalars $(\tilde c^\cA,c_\cA)$, which are related to the complex scalars $N_\cA$
via $N_\cA= i c_\cA + f_{\cA\cB} \tilde c^\cA$, but we will work directly with $N_\cA$. 
The basis form and corresponding fields are summarized in table~\ref{MBasisForms}.

\begin{table}[h!]
	\begin{center}
	\begin{tabular}{cccl}
	\toprule 
	cohomology group  \qquad& basis element &  \qquad fields \qquad \qquad& index range  \\
	\midrule
	$H^{1,1}\, (Y_4)$ & $\omega_\S $ &    $(L^\S,A^\S)$ &  $\S = 1, \ldots h^{1,1}\, (Y_4) $ \\
	$H^{3,1}\, (Y_4)$ & $\chi_{\mathcal{K}}$ & $z^{\mathcal{K}}$ & $\mathcal{K} = 1, \ldots h^{3,1}\, (Y_4) $  \\	
	$H^{2,1}\, (Y_4)$  & $\Psi^{\mathcal{A} }$  & $N_{\mathcal{A}}$ & $\mathcal{A} = 1, \ldots h^{2,1} \, (Y_4)$ \\
	$H^{3}\, (Y_4)$  & $(\alpha_\cA,\beta^\cA )$  & $(\tilde c^\cA, c_\cA)$ & $\mathcal{A} = 1, \ldots h^{2,1} \, (Y_4)$ \\
		\bottomrule
	\end{tabular}
		\caption{Relevant cohomology groups in the reduction on $Y_4$. The dimensions 
			are denoted by $h^{p,q}(Y_4) = \text{dim}\, H^{p,q}(Y_4)$. While $\Psi^\cA$ and $\chi_\cK$ are 
			complex basis elements, the forms $\omega_\Sigma$ and $(\alpha_\cA,\beta^\cA )$ constitute real basis elements.} \label{MBasisForms}
	\end{center}
\end{table}

Substituting the ansatz \eqref{JandO} and \eqref{3dAnsatz} into the action \eqref{11dAction} and 
performing a Weyl rescaling, which brings the effective action into the Einstein frame, 
we find that the three-dimensional effective theory is given by
\begin{equation}
\begin{aligned} \label{actionexp}
S^{(3)}=\int  &\frac12 R \star1-G_{\cK\bar{\cL}}\dd z^\cK\w\star\dd \bar z^{\bar{\cL}}-G_{\S\L}\dd L^\S\w\star\dd L^\L-G_{\S\L}\dd A^\S\w\star\dd A^\L  \\
& \ \quad-\frac 12 \, L^{\S}d_\S{}^{\cA\cB} \, DN_\cA \w \star D \wb N_\cB - \frac{1}{4i} \, d_\S{}^{\cA\cB} F^\S \wedge (N_\cA D \wb N_\cB - \wb N_\cB D N_\cA)\,. 
\end{aligned}
\end{equation}
Let us introduce the different objects that appear in this expression. We introduced the rescaled K\"ahler moduli
\begin{equation}
L^\S=\frac{v^\S}{\hat\cV}\,, \quad \qquad {\mathcal V}=\frac{1}{4!}\mathcal K_{\S\L\Gamma\Delta}L^\S L^\L L^\Gamma L^\Delta  ,\quad \qquad \hat\cV=\frac{1}{4!}\int_{Y_4} J^4\,,
\end{equation}
with
\begin{equation}\label{int_numbers}
\mathcal K_{\S\L\Gamma\Delta}=\int_{Y_4} \omega_\S\w\omega_\L\w\omega_\Gamma\w\omega_\Delta\,,
\end{equation}
the intersection number of two-forms. The kinetic term for the complex structure moduli $z^\cK$ depends on a K\"ahler metric  given by
\begin{equation}
G_{\cK\bar{\cL}}=-\frac{\int_{Y_4}\chi_\cK\w \chi_{\bar{\cL}}}{\int_{Y_4} \Omega\w \bar\Omega}=-\partial_{z^\cK}\partial_{\bar z^{\bar\cL}} \log \left ( \int_{Y_4}\Omega\w \bar\Omega \right ) \,,
\end{equation}
where $\chi_\cL$ are a basis of harmonic $(3,1)$-forms, with $\cL=1,\dots, h^{1,3}(Y_4)$. Regarding the kinetic terms for the vector multiplets $(L^\S,\,A^\S)$, we have that 
\begin{equation}
G_{\S\L}=\frac{\hat\cV}{4}\int_{Y_4} \omega_\S\w\star\,\omega_\L= -\frac{1}{8{\mathcal V}}\left ( \mathcal K_{\S\L}-\frac{1}{18 {\mathcal V}}\mathcal K_\S\mathcal K_\L\right ) =-\frac{1}{4}\partial_{L^\S}\partial_{L^\L}\log {\mathcal V} \,,
\end{equation}
where we defined
\begin{equation}
 \mathcal K_{\S}=\mathcal K_{\S\L\Gamma\Delta}L^\L L^\Gamma L^\Delta\,,\qquad    \mathcal K_{\S\L}=\mathcal K_{\S\L\Gamma\Delta}L^\Gamma L^\Delta\,, 
\end{equation}
and used that 
\begin{equation}
\star\omega_\S=-\frac12 J\w J \w\omega_\S+\frac{\hat\cV^2}{36}\mathcal K_{\S}J\w J\w J\,.
\end{equation}
Finally, we introduced the couplings
\begin{equation}
\int_{Y_4} \Psi^\cA\w\star\bar \Psi^\cB=L^\S d_\S{}^{\cA\cB}\,,\quad\qquad  d_\S{}^{\cA\cB} = i\int_{Y_4} \omega_\S\w \Psi^\cA\w \bar\Psi^\cB \, ,
\end{equation}
where we used that $\star\Psi^\cA=-iJ\w\Psi^\cA$. These can be written as
\begin{equation}
 \label{defQ}
d_\S{}^{\cA\cB}=-\frac12\rf^{\cB\cC}Q_{\S \cC}{}^\cA\,,\qquad \quad Q_{\S \cC}{}^\cA=M_{\S \cC}{}^\cA+i f_{\cC\cB}M_\S{}^{\cB\cA}\, ,
\end{equation}
when using the intersection numbers
\beq \label{def-Ms}
  M_{\S \cA}{}^\cB=\int_{Y_4} \omega_\S\w\a_\cA\w\b^\cB\,, \qquad M_{\S}{}^{\cA\cB}=\int_{Y_4} \omega_\S\w\b^\cA\w\b^\cB\,,
\eeq
which are independent of the K\"ahler and complex structure moduli. 
Notice that there are two important properties of the $\Psi^\cA$ that we have used numerously throughout the derivation:
\beq\label{identitiesD}
  d_\Sigma{}^{\cA\cB}=\widebar{d_\Sigma{}^{\cB\cA}}\, , \qquad  \int_{Y_4} \omega_\Sigma \wedge \Psi^\cA \wedge \Psi^\cB = 0 \, ,
\eeq
The first relation implies that $\re f^{\cA\cB}\re f_{\cC\cD}\widebar{Q_{\Sigma \cB}{}^{\cD}}=Q_{\Sigma \cC}{}^\cA$ and is the origin of the identity \eqref{identitymix}. The second identity allows to remove the intersection numbers involving  $\alpha_\cA \wedge \alpha_\cB$, such 
that the result only depends on $M_{\S \cA}{}^\cB$ and $M_{\S}{}^{\cA\cB}$ defined in \eqref{def-Ms}.

\subsection{The three-dimensional $\cN=2$ action and its symmetries}

Before manipulating the three-dimensional effective theory \eqref{actionexp} further, it is important 
to stress that it can be written in an $\mathcal N=2$ form with three-dimensional 
Yang-Mills terms~\cite{Berg:2002es}. This implies that all couplings 
are determined by a real function $K$, which we will call the kinetic potential.
Explicitly the bosonic part of the $\cN=2$ action takes the form
\begin{equation}
\begin{aligned}
\label{action_start}
S^{(3)}_{\mathcal N=2} = \int \frac{1}{2} & R \star 1 -  \wt K^{ \hat A {\bar{\hat B}}} \dd \phi_{{\hat A}} \ws \dd \wb  \phi_{{\bar{\hat B}}} 
\\ & \quad +  \frac 14 \wt K_{\S \L} \left(\dd L^\S \ws \dd L^\L + F^\S \ws F^\L\right) + F^\S \wedge \im (\wt K_{\S}^{\hat A}  \dd \phi_{\hat A})\, , 
\end{aligned}
\end{equation}
where $\phi_{\hat A}$ denotes the different complex scalar multiplets, $z^\cK$ and $N_\cA$, and $(L^\S,\,A^\S)$ corresponds to vector multiplets. 
Comparing \eqref{actionexp} with \eqref{action_start} one infers that the kinetic potential is given by
\begin{equation}
\label{K_start}
\wt K(N_\cA,z^\cK|L^\S)= - \log\Big ( \int_{Y_4} \Omega \wedge \wb \Omega\Big ) + \log \cV+ L^\S\re N_\cA\re [d_\S{}^{\cB\cA}  N_\cB] \,.
\end{equation}
It is worth pointing out that \eqref{K_start} is valid without any further assumptions about the 
real three-forms $(\alpha_\cA,\beta^\cB)$ appearing in \eqref{Psi_ansatz}.\footnote{In~\cite{Greiner:2015mdm} it was 
assumed that a basis can be chosen such that $\beta^\cA \wedge \beta^\cB = 0$ in cohomology. While this simplifies 
the computations significantly and is compatible with the weak coupling limit, it needs not necessarily be imposed in general.} 

Let us briefly discuss the symmetries of the effective action. First of all, it has an Abelian gauge symmetry given by
\begin{equation}\label{sim1}
\delta A^\S = \dd \Lambda^\S\,,
\end{equation}
where $\Lambda^\S$ is an arbitrary function. Furthermore, as advanced earlier, it has a global Abelian symmetry acting on the scalars $N_\cA$ as
\begin{equation}\label{sim2}
\delta N_\cA= i\lambda_\cA+ f_{\cA\cB}\tilde \lambda^\cB\,,
\end{equation}
with $\lambda_\cA$ and $\tilde \lambda^\cA$ real constants. These symmetries descend from large gauge transformations of the $\hat C$-field, namely $\delta\hat C = \tilde \lambda^\cA\alpha_\cA + \lambda_\cA\beta^\cA $ with $\tilde \lambda^\cA,\,\lambda_\cA\in \mathbb Z$. As usual, the classical supergravity analysis is invariant under a continuous version of the symmetry, while quantum effects break it to the discrete group. 
Using this discrete version one identifies the scalars $N_\cA$ to parameterize a complex torus
\beq \label{TM-theory}
     \mathbb{T}_{\rm M}^{2 h^{2,1}(Y_4)} = \frac{H^{2,1}(Y_4)}{ H^3(Y_4,\bbZ)}\, ,
\eeq
with a complex structure encoded by the function $f_{\cA\cB}$. Since $f_{\cA\cB}$ and $N_\cA$ vary with $z^\cK$, this torus is non-trivially fibered over the complex structure moduli space. This is reminiscent of the complex tori discussed in \eqref{complex-tori}, since one of the $z^\cK$ of the Calabi-Yau fourfold will translate to the $\tau$ in the orientifold limit. However, the three-dimensional action \eqref{action_start} with \eqref{K_start} is not yet in 
the correct duality frame in order to make the connection with the four-dimensional F-theory setting manifest. 

We will turn to the dualization and the match with a four-dimensional theory in the next subsection. Before doing this, let us point out another interesting feature 
of the above formulation. It is not difficult to check that the kinetic potential \eqref{K_start} is not invariant under \eqref{sim2}, 
but rather transforms as 
\beq \label{transK}
\delta \wt K= -\frac12 L^\S \re \Big[Q_{\Sigma}{}_\cA{}^\cB (N_\cB + \delta N_\cB) \tilde \l^\cA +(M_{\S \cA}{}^\cB \tilde \l^\cA + M_\S{}^{\cA\cB} \l_\cA)N_\cB \Big] \,.
\eeq
However, we check that this transformation yields a boundary term in the action and 
can therefore be neglected. The reason for this fact is that, in general,
the kinetic potential in \eqref{action_start} is unique up to 
\beq \label{transK_2}
    \delta\wt K=\re g(\phi) + L^\S\re h_\S(\phi) \, , 
\eeq 
where $g(\phi), h_\S(\phi)$ are 
holomorphic functions of $\phi_{\hat A}$. Indeed, using that $f_{\cA\cB}$ is holomorphic in $z^\cK$ this is precisely 
what happens in \eqref{transK}. While being in three dimensions, we have thus found a natural set of 
holomorphic functions in our setting. As we will see later, these play a key role in the up-lift to 
four dimensions and indeed reappear in the holomorphic gauge coupling function.

\subsection{Dualization of  fields to the F-theory frame}

The previous reduction is valid for any smooth Calabi-Yau fourfold. 
In order to have an F-theory background, we have to restrict to the cases in 
which $Y_4$ is elliptically fibered, which imposes certain conditions on the geometric data. 
In turn, these translate into restrictions on the three-dimensional effective action that
 ensure that it comes from the compactification of a four-dimensional theory on 
a circle. This is expected from the M-theory to 
F-theory duality and the main tool to infer information about 
F-theory effective actions. However, performing the $Y_4$ reduction as 
in subsection~\ref{smoothreduction} the resulting three-dimensional theory is generally not in 
the correct duality frame to lift it to a four-dimensional theory, so a Hodge 
star duality is usually required. Before going into the details of the dualization, 
let us illustrate this with an example.

Consider a massless chiral multiplet $\hat \Phi$ and a massless vector multiplet $\hat A$ of a four-dimensional $\mathcal N=1$ supersymmetric theory that cannot be dualized into each other. When we dimensionally reduce on a circle, we find that the chiral multiplet gives an $\mathcal N=2$ chiral multiplet $\Phi$ in three dimensions and the vector $\hat A$ yields an $\mathcal N=2$ vector multiplet, that consists of a three-dimensional vector field $A$ together with a real scalar $a$. Since the vector field $A$ is massless, it can be dualized to a real scalar $\tilde a$ which, together with $a$, corresponds to a chiral multiplet $\Phi_A$. Conversely, we can also dualize the chiral multiplet $\Phi$ into a vector multiplet
if it appears in the three-dimensional action with a real continuous shift symmetry. In general, after performing such a dualization, we can no longer lift the theory back to four dimensions. Thus, if we start with an $\mathcal N=2$ three-dimensional theory (with massless scalars and vectors) and wish to lift it to four dimensions, we first have to make sure we are in the correct duality frame.

In our case, the structure of the elliptic fibration, together with the expectations from Type IIB compactifications, is enough to find the correct frame. Following~\cite{Denef:2008wq,Grimm:2010ks}, 
we split the three- and two-forms as
\begin{equation}\label{split1}
\Psi^\cA=(\Psi^A,\Psi^\k)\, ,\qquad\quad \o_\S=(\o_\ih,\o_\a)\, ,
\end{equation}
where $\Psi^\k$ correspond to three-forms on the base of the fibration and $\Psi^A$ have components on the fiber. Similarly, the two-forms $\o_\a$, which are dual to vertical divisors, come from the base whereas $\o_\ih$ do not. In particular, the latter can be further split as
\begin{equation}\label{split2}
\o_\ih=(\o_0,\o_i)\, ,
\end{equation}
where $\o_0$ is dual to the base and $\o_i$ include the exceptional divisors and the extra sections. We 
can give a rough characterization of these forms by counting how many `legs' their components 
have in the elliptic fiber. In fact, $\o_\a$, $\Psi^\k$ have no legs in the elliptic fiber.
$\Psi^A,\o_i$ have generically components with one and zero legs
in the elliptic fiber, while $\omega_0$ has generically components with two, one and zero legs in the elliptic fiber. 
In order to have a non-vanishing coupling depending on an $Y_4$-integral over the above forms, one 
has to have a wedge-product of forms that admits at least some components with two legs along the elliptic 
fiber. One thus immediately finds  the vanishing conditions 
\beq \label{ellipticconst1}
   \cK_{\a \b \g \d }= 0\, , \qquad \cK_{i \a \b \g} = 0\, , \qquad M_{\a \k}{}^\cA = M_{\a \cA}{}^\k=M_{\a}{}^{\k \cA}=
   M_{i}{}^{\k\lambda}=0 \, ,
\eeq
The intersections $M_{0}{}^{\k \l }$ and $M_{0\k}{}^\lambda$ are in general non-vanishing. However, we can always chose 
a special three-form basis $(\alpha_\kappa,\beta^\kappa)$ such that 
\beq \label{ellipticconst2}
M_{0}{}^{\k\lambda}=0\, ,\qquad M_{0\k}{}^\lambda=\delta_\k^\lambda\, ,
\eeq
The split of the forms induces a split of the different fields as follows
\begin{equation}
N_\cA=(N_A,N_\k),\quad\quad  L^\S=(L^\ih,L^\a),\quad\quad  A^\S=(A^\ih,A^\a)\,.
\end{equation}
On the one hand, the complex fields $N_\k$ lift to a four-dimensional vectors $A^\k$ (R-R vectors) and so have to be dualized. On the other hand, the scalars $N_A$ correspond to both the $G^a$ moduli and 7-brane Wilson lines, so they remain as scalars. Regarding the three-dimensional vector multiplets, the $(A^\a, L^\a)$ lift to the four-dimensional complex scalars $T_\a$, so $A^\a$ should be dualized into a scalar. Finally, the vectors $(A^\ih, L^\ih)$ include the 7-brane vectors as well as the Kaluza-Klein vector coming from the reduction of the metric, so they are not dualized.

We are now ready to perform the dualization that brings the action \eqref{action_start} into the appropriate frame to lift to four dimensions. As usual, this can be done in a manifestly supersymmetric way by performing a Legendre transform of the kinetic potential $\wt K$ (see appendix~\ref{ap:dual} for a detailed discussion). In order to dualize the scalars $N_\k$ into vectors, we need to make sure that the kinetic potential does not depend on $\im N_\k$. At first, this is not the case for $\wt K$ given in \eqref{K_start}. 
However, we may remove such a dependence by performing a  transformation of the form \eqref{transK_2}, which yields
\begin{equation}
\begin{aligned}
\label{originalK}
\wt K= - \log \Big ( \int_{Y_4} \Omega & \wedge \wb \Omega \Big ) + \log {\mathcal{V}}+ L^\S \re d_\S{}^{\cA\cB} \re N_\cA \re N_\cB \\
& + L^\S (2 \im d_\S{}^{\k A} \re N_\k \im N_A + \im d_\S{}^{AB} \re N_A \im N_B)\,.
\end{aligned}
\end{equation}
We denote the dual kinetic potential by $K (N_A, T_\a,z^\cK|L^\ih, n^\kappa) $ and is given by
\begin{equation}
\label{Kprime}
K (N_A, T_\a,z^\cK|L^\ih, n^\kappa) = \wt K(N_A,N_\k,z^\cK| L^\ih, L^\a) -  \,  L^\a \, \re T_\a  -  \, \re N_\k \, n^\k\,,
\end{equation}
where the new variables are defined as
\begin{equation}
\label{newvar}
 \re T_\a \equiv \frac{\partial \wt K}{\partial L^\a} \, , \qquad n^\k \equiv \frac{\partial \wt K}{ \partial {\re N_\k}}  \, ,
\end{equation}
The dualized action can then be derived by inserting \eqref{Kprime} and \eqref{newvar} into the
general action \eqref{action_start}. Notice that $\re N_\k$ and $L^\a$ in \eqref{Kprime} should be understood as functions of $L^\ih, N_A, \re T_\a$, and $ n^\k$. This requires inverting the maps \eqref{newvar}, which can be done explicitly for $\re N_\k$. We find the identify 
\begin{equation}
\re N_\k = \re d_{\k\lambda}\left ( \frac12 n^\lambda - L^{\ih}\re \left [ d_{\ih}{}^{A \lambda } N_A\right  ]  \right )\,,
\end{equation}
where $\re d_{\lambda\kappa}$ is defined as the inverse of $L^{\ih}\re d_{\ih}{}^{\lambda\kappa}$. For the complex scalars $T_\a$ we only
find an implicit expression given by 
\begin{equation}
\re T_\a= \frac{\partial }{\partial L^\a}\log \cV + \re[d_{\a}{}^{AB} N_A]\re N_B\, ,
\end{equation}
This implicit form of the coordinates and kinetic potential is familiar already from the orientifold setting \eqref{Set3} and \eqref{IIBKpot}. 
However, it should be stressed that the M-theory result is more involved, since it contains the scalars $L^\ih,n^\kappa$ 
such that $K$ is not a K\"ahler potential.

Determining the dual Lagrangian is technically involved but straightforward. 
In order to do that, we have to compute the derivatives of $K (N_A, T_\a,z^\cK|L^\ih, n^\kappa) $ and express them 
in terms of derivatives of the original kinetic potential $\wt K (N_A,N_\k,z^l|L^\ih, L^\a)$. The 
details of this computation are summarized in  appendix~\ref{ap:dual}. 

\subsection{Symmetries of the dual Lagrangian}\label{sec:dualsymmetries}

Before we continue analyzing the three-dimensional Lagrangian, let us first discuss the symmetries of the dual Lagrangian. For the original Lagrangian, we found a set of Abelian symmetries given by \eqref{sim1} and \eqref{sim2}, so one might think that the symmetries of the dual Lagrangian are also Abelian. However, this is not the case~\cite{Grimm:2015ona}, which can be traced back to the existence of a Chern-Simons term in the eleven-dimensional supergravity action. In the democratic formulation we find that, due to the Chern-Simons term, the large gauge transformations of the three-form and the dual six-form potentials are not independent, but rather given by
\begin{equation}
\delta\hat C_3 = \omega_3\,,\qquad \quad \delta \hat C_6=\omega_6-\frac{1}{2}\omega_3\wedge \hat C_3\,,
\end{equation}
	with $\omega_3$ and $\omega_6$ integral closed forms. Upon dimensional reduction of the democratic action, one can check that the symmetries may be Abelian or non-Abelian, depending on how one eliminates the redundant degrees of freedom. A detailed field theory analysis in arbitrary dimension of this fact can be found in~\cite{Grimm:2015ska}.

Explicitly we can investigate the symmetries of the dual Lagrangian by translating 
the ones \eqref{sim1} and \eqref{sim2} from the original one into this new frame. In addition, one 
directly checks the new symmetries of the vectors $A^\k$ by using \eqref{action_start} with \eqref{Kprime}
and shows perfect match with the symmetries of $n^\k$ as expected by supersymmetry.
The set of gauge and global symmetries is then found to be
\begin{align}\label{dual_sym}
\begin{split}
\delta N_A &= i\lambda_A+f_{A\cA}\tilde \lambda^\cA\, , \\
\delta T_\a & = i \lambda_\a -\frac{1}{2} \tilde \lambda^\cA Q_{\a \cA}{}^B(N_B+\delta N_B)-\frac{1}{2}(M_{\a}{}^{AB}\lambda_A+M_{\a \cA}{}^B\tilde \lambda^\cA)N_B \, , \\
\delta n^\k &=- L^{\ih}(M_{\ih \cA}{}^{\k} \tilde \lambda^\cA +M_{\ih}{}^{A\k}\lambda_A)\, , \\
\delta A^\k &= \dd\Lambda^\k -  A^\ih (M_{\ih \cA}{}^{\k} \tilde \lambda^\cA +M_{\ih}{}^{A\k}\lambda_A)\, ,  \\
\delta A^\ih & = \dd\Lambda^\ih\, , 
\end{split}
\end{align}
where $\lambda_\a$, $\lambda_A$, $\tilde \lambda^\cA$ are arbitrary real constants and $\Lambda^\k$, $\Lambda^\ih$ are arbitrary real functions. Notice that the right hand side of $\delta N_A,\,\delta T_\a$ is holomorphic and that the transformation is valid for finite values of $\lambda_\a$, $\lambda_A$, and $\tilde \lambda^\cA$ .

The symmetry group is now non-Abelian and, in particular, it is a generalization of the Heisenberg group. Notice also that, unlike for the original Lagrangian, the symmetries of the scalars and vectors are mixed. This can be seen from the transformation rule for $A^\k$, that depends on $\tilde \lambda^\cA$ and $\lambda_A$, inducing a \emph{constant} change of basis in the space of $U(1)$'s (see also~\cite{Marchesano:2014bia}). This necessarily implies that the gauge coupling function must depend on the scalars and transform under the symmetries appropriately in order to make the whole Lagrangian invariant.
Furthermore, if we were to gauge the global (non-Abelian) symmetry by promoting $\tilde \lambda^\cA$ and $\lambda_A$ to be arbitrary functions, we find that the transformation of the vectors is no longer constant and precisely matches that of a non-Abelian vector field~\cite{Grimm:2015ona}.

More explicitly, in order that the three-dimensional kinetic terms are invariant under \eqref{dual_sym}, 
i.e.
\beq
   \delta \left ( K_{\cI\cJ}F^\cI \ws F^\cJ \right )= 0 \, , 
\eeq
the three-dimensional gauge kinetic terms should transform, for finite $\tilde \lambda^\cA=(\tilde \lambda^A,\tilde \lambda^\kappa)$ 
and $\lambda_A$, as
\begin{equation}
K_{\cI \cJ}\,\longrightarrow\, \m_{\cI}^\cK\,\m_\cJ^\cL\, K_{\cK\cL}\,,
\end{equation}
with 
\begin{equation}\label{transm}
\m_\cI^\cJ=\left (\begin{array}{cc}
\delta_\l^\k& \qquad M_{\jh \cA}{}^{\k} \tilde \lambda^\cA +M_{\jh}{}^{A\k}\lambda_A\\
0&\quad \delta_\jh^\ih\end{array}\right )\,.
\end{equation}
Here the indices $\cI,\,\cJ,\dots$ run over all the three-dimensional vectors, namely $A^\k$ and $A^\ih$.

As we will see in the next section, the couplings $M_{i \cA}{}^{\k}$ and $M_{j}{}^{A\k}$
are related to kinetic mixing of 7-brane and bulk gauge fields, while $M_{0 I}{}^{\k}$ and $M_{0}{}^{A\k}$
have no immediate four-dimensional meaning. We would like to stress at this point that, in three dimensions, 
the coefficient of the kinetic terms of the vectors $K_{\cI\cJ}$ is invariant if and only if the kinetic mixing is zero and 
$M_{0 \cA}{}^{\k}$ and $M_{0}{}^{A\k}$ vanish. This carries over to a property of the 
four-dimensional gauge coupling function, as we show in the following.

\section{Determining F-theory gauge coupling functions}\label{sec:four}

Having determined the three-dimensional action in the correct duality frame, we can compare it with the circle reduction of an arbitrary four-dimensional action.
As shown in appendix~\ref{ap:circle}, the circle reduction of a four-dimensional $\mathcal N=1$ supergravity action \eqref{act4D}
yields a three-dimensional $\mathcal N=2$ supergravity given by \eqref{action_start}, with kinetic potential 
\begin{equation}\label{K3dred}
K(M|R,\xi) = \hat K(M,\widebar M) + \log R - \frac{1}{2R} \re \hat f_{\mathbf{I}\mathbf{J}}(M) \, \xi^{\mathbf{I}} \xi^{\mathbf{J}} \, . 
\end{equation}
Here we set $R=r^{-2}$, with $r$ being the radius of the circle, and introduced the scalars $\xi^{\mathbf{I}}$ that come from reducing the four-dimensional vector fields. The index $\mathbf{I}$ runs over the four-dimensional vector fields and is split as $\{\k,i\}$. From now on, we denote four-dimensional quantities by a hat.

\subsection{Transformation rules of the gauge coupling functions}

Before we proceed to compare the result obtained from the (dualized) M-theory reduction with a generic four-dimensional theory on a circle, let us
discuss the transformation properties of the four-dimensional gauge coupling functions. In the last section we saw that, in general, the kinetic terms of the
three-dimensional vectors $K_{\cI\cJ}$ transform under the shift-symmetries of the scalars. Clearly, the four dimensional gauge coupling function shares
a similar property. Indeed, consider the ansatz for a four-dimensional vector on a circle, namely
\beq\label{ansatz4dv}
\hat A^{\mathbf{I}}=A^{\mathbf{I}}-\frac{\xi^{\mathbf{I}}}{R}(\dd y+A^0) \,,
\eeq
where $\dd y$ is the non-trivial one-form on the circle. We also introduced the Kaluza-Klein vector $A^0$ coming from the reduction of the metric on a circle, namely
\beq
\dd\hat s^2=\dd s^2+\frac{1}{R}(\dd y+A^0)^2\,.
\eeq
Using \eqref{dual_sym} together with \eqref{ansatz4dv}, we find that the transformation of the four-dimensional vector on $\mathbb R^{1,2}\times S^1$ is
\begin{subequations}
\begin{align}
\delta \hat A^\k&=\dd\Lam^\k -A^i(M_{i A}{}^{\k} \tilde \lambda^A +M_{i}{}^{A\k}\lambda_A) + (M_{0I}{}^\k\tilde\l^I+M_0{}^{Ak}\lam_A)\dd y \label{trans_vector4d} \,, \\
\delta \hat A^i&=\dd\Lam^i \,,
\end{align}
\end{subequations}
where we used that $L^0=R$ and $\frac{L^i}{R}\rightarrow 0$. Since $\dd y$ is the non-trivial one-form on $S^1$, we recognize the last term in \eqref{trans_vector4d} as a large gauge transformation. These transformations along the circle are often key in investigating the properties of the
F-theory effective action as recently demonstrated in~\cite{Grimm:2015zea,Grimm:2015wda} for 7-brane gauge fields. Here we find a non-trivial
completion of these transformations  to include R-R bulk gauge fields. 
In the decompactification limit, large gauge transformations are meaningless since there are no
non-trivial one-forms in $\mathbb R^{1,3}$. Thus, we find that the transformation of the vectors in $\mathbb R^{1,3}$ is
\begin{align}\label{trans_vector4db}
\begin{split}
\delta \hat A^\k&=\dd\hat \Lam^\k -A^i(M_{i A}{}^{\k} \tilde \lambda^A +M_{i}{}^{A\k}\lambda_A) \,, \\
\delta \hat A^i&=\dd\hat \Lam^i\,,
\end{split}
\end{align}
where now $\hat\Lam^\k$ and $\hat\Lam^i$ are arbitrary functions in four-dimensions. This shows that, under shifts of the four-dimensional scalars $\hat N_A$, the vectors $\hat A^\k$ transform non-trivially only when $M_{iA}{}^{\k}$ or $M_{i}{}^{A\k}$ are different from zero. This is the M-theory derivation of the result given in eq.~\eqref{transAIIB}.

Now from \eqref{trans_vector4db} and \eqref{act4D}, we can readily determine the transformation rules for the four-dimensional gauge coupling function $\hat f_{\mathbf{I}\mathbf{J}}$, namely
\begin{equation}
 \hat f_{\mathbf{I}\mathbf{J}}\,\longrightarrow\, \hat m_{\mathbf{I}}^{\mathbf{K}}\,\hat m_{\mathbf{J}}^{\mathbf{L}}\, \hat f_{\mathbf{K}\mathbf{L}} + iC_{\mathbf{I}\mathbf{J}} \,,
\end{equation}
with 
\begin{equation}\label{transm2}
\hat m_{\mathbf{I}}^{\mathbf{J}}=\left (\begin{array}{cc}
\delta_\l^\k& \qquad M_{j A}{}^{\k} \tilde \lambda^A +M_{j}{}^{A\k}\lambda_A\\
0&\quad \delta_j^i\end{array}\right )\,,
\end{equation}
We included the possibility of having a constant shift $C_{\mathbf{I}\mathbf{J}}$ in $\im \hat f_{\mathbf{I}\mathbf{J}}$. Splitting the indices, this corresponds to
\begin{align*}
\delta \hat f_{\k\l} &=i C_{\k\l}\\
\delta\hat f_{i\k} &=p_{i}^\k\hat f_{\k\l}+iC_{\k\l}\\
\delta\hat f_{ij} &=p^\k_{i}\hat f_{j \k}+p^\k_{j}\hat f_{i \k} + p_i^\k p_j^\l\hat f_{\k\l}+i C_{ij} \,,
\end{align*}
with
\begin{equation}
p_{i}^\k=M_{i A}{}^{\k} \tilde \lambda^A +M_{i}{}^{A\k}\lambda_A\,.
\end{equation}

Finally, notice that when $M_{i A}{}^\k= M_{i}{}^{A\k}=0$, we find that the gauge coupling function must be invariant, up to possibly constant shifts of its imaginary part. In the following we will see that this corresponds to the case in which the kinetic mixing between the four-dimensional vectors $\hat A^\k$ and $\hat A^i$ vanishes.

\subsection{Gauge coupling functions from dimensional reduction}

In the following, we compare the action derived from the kinetic potential \eqref{Kprime} with the one derived from \eqref{K3dred}, paying special attention to the gauge coupling function. In order to do so, we will need the derivatives of the dual kinetic potential $K (N_A, T_\a,z^\cK|L^\ih, n^\kappa)$, which are given in appendix~\ref{ap:dual}.

\subsubsection{On the weak string-coupling limit}

In addition to presenting the F-theory result we will also study the restriction 
to the weak string-coupling limit discussed in section~\ref{sec_TypeIIB}.   
In order to do that it is useful to point out the matching of the moduli. 
First, note that the complex structure moduli $z^{\cK}$
of $Y_4$ correspond to the complex structure moduli of the double cover $Y_3$ of $B_3$, 
the axio-dilaton $\tau$, and the D7-brane deformations $\zeta_K$:
\beq \label{zksplit}
   z^\cK \quad \xrightarrow{\quad \rm{weak\ coupl.}\quad } \quad z^k, \tau, \zeta_K\, , 
\eeq 
which are the fields in the Set 1 given in \eqref{Set1}.\footnote{Note that we have not included $\zeta_K$ in 
the orientifold analysis. In F-theory a general $z^\cK$-dependence automatically includes these moduli.} 
Second, the F-theory moduli $N_A$ are naturally split as 
\beq \label{Nsplit}
   N_A \quad \xrightarrow{\quad \rm{weak\ coupl.}\quad } \quad (a_p , G^a)\ ,
\eeq
where $a_p$ are the D7-brane Wilson line moduli and $G^a$ are the 
R-R and NS-NS two-form moduli constituting the Set 2 given in \eqref{Set2}. 
Third, recalling the result \eqref{RR-gaugecoupling} and the definitions \eqref{AD7_exp2}, \eqref{Set2} we 
note that one identifies\footnote{The identification of $f_{\k\l}$ with \eqref{RR-gaugecoupling} will become 
apparent in the next paragraphs.}
\beq
  \left( \begin{array}{cc} \ f_{\kappa \lambda}\ & \ f_{\kappa B}\ \\ \ f_{A\lambda}\ & \ f_{AB}\ \end{array} \right) \quad \xrightarrow{\quad \rm{weak\ coupl.}\quad }\quad 
  \left( \begin{array}{cc} -\cF_{\kappa \lambda} |_{z^k=0} &\ 0\ \\ 
  \ 0 \ &\ (f_{pq} ,  - i \tau \delta^{ab})\end{array}\right)  \, , \\
\eeq
where we stress that $\cF_{\kappa \lambda}$ and $f_{pq}$ are only determined as 
functions of the complex structure moduli of $Y_3$. The F-theory result is 
significantly more general, since it encodes the full dependence on all 
complex structure moduli $z^\cK$ of $Y_4$.  Applying the split \eqref{zksplit}
it can be used to derive corrections to the orientifold result.

\subsubsection{Gauge coupling function for R-R vectors}
Let us start with the derivation of the four-dimensional gauge coupling function for the R-R vectors, namely $\hat f_{\k\l}$. From the results in appendix~\ref{ap:circle}, we immediately see that the real part of the gauge coupling function is encoded in $K_{\k \lambda}$, which is the kinetic term for the three-dimensional vectors $A^\k$. According to eq.~\eqref{app1}, it is given by
\begin{equation}\label{RRcoupling}
K_{\k \lambda} =  \frac 1{R} \re f_{\k \lambda}\,,
\end{equation}
where we assumed that
\begin{equation} \label{f-assumption}
f_{\k A}=0\,,\qquad\quad L^0=R\,.
\end{equation}
These assumptions appear to be essential. They greatly simplify the results and, in particular, they make \eqref{RRcoupling} into the real part of a holomorphic function, which matches the expectations from the Type IIB perspective. Thus, we will assert that \eqref{f-assumption} holds for the rest of the paper.  It would be interesting to show that the vanishing condition $f_{\k A}=0$ can be proved for elliptic fibrations.

The computation of the imaginary part of the four-dimensional gauge coupling function is a bit more involved. However, by carefully tracking the
circle reduction, we see that it is encoded in the three-dimensional action in the couplings 
\beq
F^\k \wedge \im ( K_{\k}^{\hat A}  \dd \phi_{\hat A})
\eeq
in \eqref{action_start}, where $\hat A$ runs over all the chiral fields in three-dimensions. According to the results in appendix~\ref{ap:dual}, we find that
\beq\label{RRcouplingim}
F^\k \wedge \im (K^{ \hat A}_\k \dd \phi_{\hat A}) = \frac{n^\lambda}{2R} F^\k \wedge \dd \im f_{\k\l} + \frac{L^i}{2R} F^\k \wedge \dd \im ( Q_{i\k}{}^a N_a).
\eeq
In particular, the imaginary part of $\hat f_{\k\l}$ is encoded in the coefficient that multiplies $n^\l/R$ above. Thus, from \eqref{RRcoupling} and \eqref{RRcouplingim}, we conclude that the
four-dimensional gauge coupling function for the R-R gauge bosons is given by
\beq \label{hatf1}
\hat f_{\k\l}=-f_{\k\l}\, ,
\eeq
which is holomorphic in the complex structure moduli of the Calabi-Yau fourfold, and therefore 
holomorphic with respect to the four-dimensional 
chiral fields. 
The result \eqref{hatf1} is in accord with the expectations from the Type IIB orientifolds, c.f.~\eqref{RR-gaugecoupling}.
However, it is important to note that the F-theory result \eqref{hatf1} is significantly more general, 
since the function $f_{\k\l}$ can depend on all complex structure moduli of $Y_4$.

\subsubsection{Kinetic mixing between R-R and 7-brane vectors}

Now we move on to considering the kinetic mixing $\hat f_{\k i}$ between the open and closed string gauge bosons. From the circle reduction, we see that $\re \hat f_{\k i}$ is encoded in $K_{\k i}$, the three-dimensional kinetic mixing between $A^\k$ and $A^i$. We find that the M-theory reduction yields
\begin{equation}\label{mixingcoupling}
K_{\k i} = \frac 1{R} \re \left[Q_{i \k }{}^A N_A \right]\,, 
\end{equation}
where $Q_{i \k }{}^A$ is the holomorphic function defined in \eqref{defQ}. Notice that \eqref{mixingcoupling} is again the real part of a holomorphic function of the complex moduli. This also shows that the mixing is proportional to the couplings $M_{i \k}{}^A$ and $M_{i}{}^{A\k}$, which are related to the ones that appear in \eqref{trans_vector4db} by the identity \eqref{identitiesD}. This proves the statement in the last section that the transformation for the vector is trivial if and only if the mixing vanishes.

Just like in the previous case, we can compute the imaginary part of the mixing $\im \hat f_{\k i}$ by analyzing \eqref{RRcouplingim}. In this case, it is given by the term proportional to $L^i/R$. Thus, we find that
\beq \label{hfkappai}
\hat f_{\k i} =- Q_{i\k}{}^A N_A= -(M_{i \k}{}^A+i f_{\k \l}M_i{}^{\l A})N_A\, ,
\eeq
which is holomorphic in both the complex structure moduli $z^\cK$ and the moduli $N_A$. 

The identification \eqref{hfkappai} agrees with the result given in section~\ref{TypeIIBmixing}, when 
asserting that $M_i{}^{\l A}$ is only non-vanishing for the directions of the Wilson line moduli $a_p$. However, let us stress again that 
in order to match it with the results obtained in~\cite{Jockers:2004yj} from dimensional reduction of the D7-brane action we had to use heavily the identities \eqref{identitiesD}, which
were not known in the Type IIB context (see the discussion around eq.~\eqref{kinmix}).

Let us briefly mention that we can compute the mixing between the Kaluza-Klein vector and the R-R vectors, which is
\begin{equation}
K_{\k 0} = -\frac{1}{R^2} \left ( n^\lambda  \re f_{\k \lambda}  + L^i \re \left[ Q_{i\k}{}^A N_A \right] \right )\,.
\end{equation}
Of course, this has no meaning in four dimensions. However, it is reassuring to check that it is what one would expect from a theory that comes from a circle reduction, given \eqref{RRcoupling} and \eqref{mixingcoupling}.

\subsubsection{Gauge coupling function for 7-brane vectors}

Finally, let us discuss the gauge coupling function $\hat f_{ij}$ for the seven-brane gauge fields that, 
as we saw in section~\ref{sec:IIBcoupling}, is the most
involved coupling. In particular, we do not expect to obtain a holomorphic gauge coupling function $\hat f_{ij}$ directly from dimensional reduction. In
the following we simply give the result that we obtain from dimensional reduction and in the next subsection we then discuss how one can use
holomorphicity and the discrete shift-symmetries of the axions to constrain the exact result.

Following the same strategy as before, we see that $\re \hat f_{ij}$ is given by $K_{ij}$, the three-dimensional kinetic terms for the 7-brane gauge bosons. There is, however, a further complication when discussing this 
coupling that has to be addressed. As shown in appendix~\ref{ap:dual}, in
terms of the original kinetic potential $\wt K$, it reads
\begin{equation}\label{open}
K_{ij} = \wt K_{ij} -  \wt K_{i \a} \wt K_{j \b} (\wt K_{\a\b})^{-1} - \wt K_i ^\k \wt K_j^\lambda (\wt K^{\k \lambda})^{-1}\,,
\end{equation} 
with $\wt K$ given by \eqref{originalK}. Thus, we immediately see that $K_{ij}$ depends on all the possible intersection numbers \eqref{int_numbers}, but we do not expect all of them to contribute to the gauge coupling function in four dimensions. In particular,
the couplings $\cK_{ijkl}$ and $\cK_{ijk\a}$ induce a dependence of $K_{ij} $ on the scalars $L^i$, which have 
no four-dimensional scalar analog.  
This suggests that, just like in~\cite{Grimm:2011fx,Cvetic:2012xn,Cvetic:2013uta}, the classical M-theory reduction contains terms that correspond to one-loop effects from the circle reduction of the four-dimensional theory. However, notice that unlike in ~\cite{Grimm:2011fx,Cvetic:2012xn,Cvetic:2013uta}, we are performing a dimensional reduction without fluxes, so the four-dimensional theory is non-chiral in our case. Thus, the smooth Calabi-Yau fourfold encodes information about non-chiral states.
We leave a more detailed study of these corrections and their interpretation for future work. 

In order to match the classical circle reduction, we will compute the coupling \eqref{open} assuming that the only non-vanishing intersection numbers are
\begin{equation}
\mathcal K_{0\a\b\g}\equiv\mathcal K_{\a\b\g}\,,\qquad \quad \mathcal K_{\a\b i j}\equiv - C^\g_{ij}\mathcal K_{\a\b\g}\,,
\end{equation}
where $\mathcal K_{\a\b\g}$ are the intersection numbers of the two-forms on the base of the elliptic fibration. We also expressed the intersection numbers $\mathcal K_{\a\b i j}$ in terms of those of the base. The precise interpretation of the divisors labeled with indices 
$i,j$ depends on the model under consideration. The first possibility is that $i,j$ are labeling 
exceptional divisors over a single non-Abelian 7-brane wrapping a divisor $S$ in the base $B_3$. In this case 
one can expand the Poincar\'e-dual two-form as $[S] = \delta_7^\a \omega_\a|_{B_3}$ and split $C^\a_{ij} =   \delta_7^\a C_{ij}$, where $C_{ij}$
is the Cartan matrix of the non-Abelian gauge algebra.\footnote{In order to have this simple identification one has to restrict to ADE gauge algebras.}
A second possibility is that the indices $i,j$ label multiple $U(1)$ gauge factors stemming from several 7-branes on different divisors in $B_3$.
In this case it is convenient to keep $C_{ij}^\alpha$ in this general form, since this allows us to include kinetic mixing among the 7-brane $U(1)$'s.
In either case, we compute to linear order in $C^\a_{ij}$ that 
\begin{equation}\label{openfunction}
K_{ij}=-\frac{3}{R\,\mathcal K^{\textbf{b}}} C_{ij}^\a\mathcal K^{\textbf{b}}_\a+\frac{1}{R}\re[d_i{}^{A\k}N_A]\re[d_j{}^{B\l}N_B]\re f_{\k\l}\,,
\end{equation}
where we defined
\begin{equation}
\mathcal K^{\textbf{b}}=\mathcal K_{\a\b\g} L^\a L^\b L^\g\,,\qquad \mathcal K^{\textbf{b}}_\a=\mathcal K_{\a\b\g} L^\b L^\g\,.
\end{equation}
In this expression, on the one hand, the first term in \eqref{openfunction} is proportional to the volumes of the divisors in $B_3$ specified by $C_{ij}^\a$. 
From the Type IIB perspective this corresponds to the fact that the gauge coupling scales with the volumes of the cycles wrapped by the 7-branes. The second term, on the other hand, is proportional to the couplings $M_{i A}{}^\k$ and $M_{i}{}^{A\k}$ and, in particular,  vanishes when there is no mixing between $A^\k$ and $A^i$. Notice that, as expected from the Type IIB discussion, \eqref{openfunction} is \emph{not} the real part of a holomorphic function of the chiral fields, even in the absence of mixing. Indeed, from \eqref{newvar} we have that
\begin{equation}\label{reta}
C_{ij} ^\a \re T_\a=\frac{3}{\mathcal K^{\textbf{b}}}   C^\a_{ij}\mathcal K^{\textbf{b}}_\a+  C^\a_{ij}\re [d_\a{}^{BA} N_B]\re N_A\,,
\end{equation}
which contains a term proportional to the square of $N_A$ that is missing in \eqref{openfunction}. This is precisely the same problem we encountered in the Type IIB setting of section~\ref{sec:IIBcoupling}, where the contribution proportional to the square of the Wilson lines does not arise from dimensional reduction.

Finally, let us mention that the second term in \eqref{openfunction} is holomorphic in $N_A$ if and only if we have that
\begin{equation}
Q_{i\k}{}^A Q_{jB}{}^\k=0\,.
\end{equation}
However, this is not sufficient to guarantee holomorphicity in complex structure moduli $z^\cK$ of $Y_4$. In the following we discuss in detail the corrections that are needed, in four dimensions, to have a holomorphic gauge coupling function. However, we focus on the case without kinetic mixing of 7-brane and R-R gauge fields and leave the general case to future work.

\subsection{Shift symmetries, quantum corrections, and theta functions}

In the previous subsection we have shown that a direct dimensional reduction 
of eleven-dimensional supergravity on a smooth Calabi-Yau fourfold yields 
vector kinetic terms and a complex moduli space that appear to be incompatible 
with a reduced four-dimensional holomorphic gauge coupling function. 
In the absence of kinetic mixing the missing terms in the completion to a holomorphic result are 
of the form $ \re (d_\a{}^{BA} N_B)\re N_A$. From our detailed discussion 
of the orientifold setting in section~\ref{sec_TypeIIB}, however, we should be alerted 
that this apparent conflict was already encountered for D7-brane 
Wilson line moduli. In fact, we recalled in subsections~\ref{sec:IIBcoupling} and~\ref{sec:theta-ori} that 
the corrections to the gauge coupling functions quadratic in the Wilson line 
moduli are only generated at one string-loop order and therefore are not 
found by a dimensional reduction of the tree-level D7-brane effective action.
We observe that in F-theory effective actions derived via eleven-dimensional 
supergravity a similar feature occurs for all moduli $N_A$, i.e.~both the Wilson line moduli 
and the R-R and NS-NS two-form moduli in the split \eqref{Nsplit}. This implies that 
to ensure holomorphicity of the 
gauge coupling function in $T_\alpha$ one needs to include in the M-theory 
reduction a quantum correction of the form
\beq \label{fquant}
   f^{\rm quant}_{ij} = C^\alpha_{ij} \, d_\a{}^{BA} N_B \re N_A + \ldots\, ,
\eeq
In the M-theory setting it is much harder to identify the origin of such 
a correction. One expects that it 
arises due to certain M2-brane states, by following the F-theory to M-theory 
duality, but it remains an open question how to make this more precise. 
As we will see in the following  we can nevertheless 
infer non-trivial constraints on $f^{\rm quant}_{ij}$ by using symmetries and the expected 
holomorphicity properties of the effective theory. For simplicity we will only discuss 
the case without kinetic mixing in the rest of this work.

In order to proceed we begin by collecting a few observations supporting the fact that important 
corrections have to be missing in the reduction of the supergravity action.
On the one hand, it is clear from the outset that the three-dimensional reduction result 
is invariant under all the shift-symmetries \eqref{dual_sym} even when choosing 
continuous parameters $\l_A$, $\l_\a$, $\tilde \l^\cA$. Since these symmetries are inherited from the  
eleven-dimensional action and unbroken throughout the classical reduction, there is 
simply no way how they could be broken. On the other hand, we have argued in subsection~\ref{sec:theta-ori} 
that in the presence of the fields $N_A$ the \textit{continuous} symmetries $\l_A$, $\tilde \l^A$ 
acting non-trivially on the holomorphic gauge coupling 
function in four dimensions are always broken. The \textit{discrete} symmetries
are, however, manifest when including 7-brane fluxes or quantum corrections
resulting in a theta function on the complex torus spanned by $N_A$. 
We expect that this is equally the case for a full-fledged F-theory compactification, such 
that indeed corrections must be missing in the above dimensional reduction.

Note that in the M-theory background \eqref{backr} we did not include any background fluxes 
$\langle d\hat C\rangle$ on $Y_4$. This implies that the F-theory setting will not 
contain background fluxes either and, in particular, we did not consider 7-branes with 
world-volume fluxes. This implies that the manifestation of the discrete symmetries 
for the $G^a$ moduli obtained in \eqref{def-cF} by completing $i^*B_2 - 2\pi \alpha' {F}$, 
requires an extension of our M-theory analysis. In fact, it was argued in~\cite{Mayrhofer:2013ara}
that such orientifold fluxes are precisely the ones that correspond to so-called 
hypercharge fluxes in F-theory GUTs~\cite{Beasley:2008kw,Donagi:2008kj}. They neither induce a D-term nor an F-term potential 
for the considered moduli, 
but nevertheless can, for example, break a non-Abelian gauge group. In our context they are 
crucial to make the discrete symmetries manifest. It is of enormous importance to 
understand the manifestation of these fluxes in the M-theory reduction in greater detail. 

The second possibility encountered in subsection~\ref{sec:theta-ori} was a manifestation of 
the discrete shift-symmetries by completing the first term in \eqref{fquant} with a theta function.
In fact, note that the $N_A$ span a complex torus $\mathbb{T}_{\rm F}^{2n}$ of real dimension
$2n = 2(h^{2,1}(Y_4) - h^{2,1}(B_3))$. Its complex structure is determined by the holomorphic 
function $f_{AB}$ and by using our assumption $f_{\l A} = 0$, as given in \eqref{f-assumption}, and 
the restriction to a setting without kinetic mixing this torus 
arises trivially in the split of $\mathbb{T}^{2h^{2,1}(Y_4)}_{\rm M}$ defined in \eqref{TM-theory}.
We then define a line bundle $\cL$ on this torus analog to the one in 
subsection~\ref{sec:theta-ori}. Freezing the complex structure moduli of $Y_4$
one defines  the  connection in holomorphic gauge
\beq \label{conno_F}
   \mathfrak{A}_{ij }^{\rm h}=\frac{i}{4} C_{ij}^\a \left (2  M_{\a A}{}^B \re f^{AC} \re N_C+ M_{\a}{}^{AB} N_A\right ) \dd N_B\, ,
\eeq
such that $\mathfrak{F}_{ij } = \dd \mathfrak{A}^{\rm h}_{ij }$ is a $(1,1)$-form. Note that this expression is still in the three-dimensional 
Coulomb branch as indicated by the indices $i,j$. While the lift with a non-Abelian gauge group is more involved, one realizes 
that for a single $U(1)$ gauge group factor one finds the generalization of \eqref{conno}. In the following we will restrict 
to this Abelian case and drop the indices $i,j$.
Arguing as in subsection~\ref{sec:theta-ori} one can use this connection in the 
non-holomorphic gauge and look for holomorphic sections 
\beq \label{PsiF}
   \Psi = \text{exp}\Big(C^\alpha \, d_\a{}^{BA} N_B \re N_A \Big)\, \Theta(N_A,z^\cK) \, , 
\eeq
Here, as  in subsection~\ref{sec:theta-ori}, $\Theta$ is a sum of Riemann theta functions with $z^\cK$-dependent 
coefficients, in general. 

Unfortunately, we do not know an M-theory argument how $\Theta$ can be fully determined. 
In addition to the ambiguities in the complex structure dependent coefficients, one also faces the 
fact that fluxes should be properly included into \eqref{PsiF}. One might speculate that 
the some of the constants $(\nu_a,\mu^a)$ determining the shifts in the theta functions \eqref{gen-theta} might admit an 
interpretation as fluxes. However, we also expect that the non-holomorphic pre-factor and hence the 
line bundle and connection become modified. It would be very interesting to investigate the 
proper inclusion of fluxes in future work.

\section{Conclusions} \label{sec:conclusions}

In this paper we have studied the gauge coupling functions arising in $\mathcal N=1$ Type IIB orientifolds with
D7-branes and F-theory. First, we have analyzed the result that one obtains from dimensional
reduction of Type IIB supergravity coupled to the D7-brane action without kinetic mixing between the open and
closed string gauge fields following~\cite{Jockers:2004yj}. We have seen that this
does not yield a gauge coupling function which is holomorphic in the chiral coordinates and, therefore, has to be modified.
As already mentioned in~\cite{Jockers:2004yj}, one expects that corrections coming from open-string one-loop effects generate precisely
the missing terms to establish a holomorphic result. However, an explicit computation of such corrections is very challenging
and has only been done in a related setting in toroidal orbifolds~\cite{Berg:2004ek}. We have shown that by carefully analyzing the
shift-symmetries of the closed string and open string axions in the effective field theory, one can severely constrain the specific structure of such corrections even in generic Calabi-Yau vacua. 

In Type IIB orientifolds we have discussed two mechanisms to ensure that the gauge coupling 
function $\hat f_{\rm D7}$ transforms appropriately under discrete shift-symmetries. On the one hand,  we reviewed the inclusion of 
D7-brane world-volume flux to make the symmetries of the R-R and NS-NS two-form moduli $G^a$ manifest.  
On the other hand, we have stressed that gauge coupling function in general also depends on the complex Wilson line
moduli $a_p$, which also admit discrete shift-symmetries. In fact, they span a complex torus with complex structure 
determined by a function $f_{pq}$, which is itself holomorphic in the complex structure moduli. 
Then, by simply imposing holomorphicity and invariance under such symmetries, 
we obtain that the required one-loop corrections are encoded by a holomorphic
section $\Psi=\text{exp}({\hat f_{\rm D7}^{\rm 1-loop}})$ of a certain line bundle defined 
on the torus spanned by the Wilson lines. Constructing the connection on this line bundle, 
such sections are then found to be comprised of 
a term quadratic in the Wilson lines, required for holomorphicity of the complete $\hat f_{\rm D7}$, 
and a sum of Riemann theta functions with, in general, complex structure dependent coefficients. 
This form of the gauge coupling function is in agreement with the results in~\cite{Berg:2004ek}, even 
though in our setting the torus is in general not related to the compactification space. 

It is important to stress that we did not  unravel the precise physical interpretation of having to deal with 
holomorphic sections $\Psi$ of the constructed line bundle. We were lead to this construction by holomorphicity 
and symmetries of the gauge coupling function, but  we were not able to completely  
fix the choice of $\Psi$ appearing in the gauge coupling function. Our construction, however, 
is reminiscent of the consideration first given in~\cite{Witten:1996hc}. In this work the 
partition function of an M5-brane is constructed and a similar ambiguity of choosing the 
correct section had to be addressed. One might hope that 
the extensions of~\cite{Witten:1996hc} to Type IIB supergravity with D-branes~\cite{Belov:2006jd,Belov:2006xj} might 
shed new light on the significance of the choice of $\Psi$ in our setting.
It is also intriguing to point out that the complex 
structure dependence of $\Psi$ might be fully constrained when 
identifying it as a wave-function of a quantum system along the lines of~\cite{Witten:1993ed}.
In would be interesting to check whether these ideas can be made more explicit for our setting. 

Extending our analysis of the Type IIB orientifold setting we have also included the effects of  kinetic mixing 
between D7-brane gauge fields and R-R gauge fields. In particular, we derived that when the mixing is
non-zero, the gauge coupling function should not be invariant under the shift-symmetries, since these induce a constant
change of basis in the space of gauge bosons, mixing open and closed string $U(1)$'s. Our systematic approach allowed us to clarify
certain puzzles that appeared in~\cite{Grimm:2015ona}. In particular, we argued that it is indeed possible to gauge 
specific non-Abelian isometries by Abelian vectors even though the gauge coupling function is independent of such gaugings. 
The underlying structure is omnipresent in string theory models and stems from the fact that higher-degree R-R form potentials 
admit non-trivial symmetry transformations under lower-degree forms from the brane or bulk theory. 
It would be interesting to see whether the ideas to exploit the stringy symmetries for the axions and gauge fields
can be generalized further.

In the second main part of the work, we have studied the gauge coupling function for genuine 
F-theory backgrounds via dimensional reduction
of M-theory on a Calabi-Yau fourfold. One of the main advantages of this approach is that many of the moduli that appear to be
completely different from the Type IIB perspective, turn out to have a common origin in the Calabi-Yau fourfold.
In addition to being also applicable away from weak string coupling, the F-theory settings also 
allow us (1) to fully include the dependence on the 7-brane position moduli, (2)  derive interesting and useful 
relations between different moduli that are obscure in the
IIB picture, and (3) provide geometric arguments for the properties of the various couplings in the bulk and 7-brane 
sector. 

In order to investigate the gauge coupling function we have crucially 
extended the results in~\cite{Grimm:2010ks}. We performed the M-theory reduction in full generality and explained
in detail the role of the
elliptic fibration when performing the dualization to the F-theory frame. 
In doing so, we have payed special attention to the shift-symmetries
of the axions coming from the M-theory three-form expanded into three-forms 
of the Calabi-Yau fourfold. We have shown explicitly that a direct reduction of 
eleven-dimensional supergravity at first only yields shift-symmetries that 
are Abelian. Due to the dualization of three-dimensional fields into the F-theory frame 
they become non-Abelian as already discussed in~\cite{Grimm:2015ona,Grimm:2015ska}. 
As we have seen, this is a direct consequence of having a
non-trivial Chern-Simons coupling in the eleven-dimensional supergravity action. Furthermore, it provides the M-theory origin
of the more involved shift-symmetries in Type IIB compactifications.

We then determined the four-dimensional gauge coupling functions of the F-theory setting, 
by comparing the three-dimensional M-theory effective action with the circle reduction of a four-dimensional theory.
As in the Type IIB orientifolds, the resulting gauge coupling function is at first not holomorphic. In fact,  
the reduction of eleven-dimensional supergravity does not capture any of the quadratic 
corrections in the $T_\alpha$ coordinates determined from the scalar kinetic terms. This 
is compatible with the fact that the dimensional reduction does not break the continuous shift-symmetries 
and indicates that important quantum corrections are missed. 
However,
by mimicking the arguments we made for the Wilson line moduli in Type IIB orientifolds, 
we derived that an appropriate correction to the F-theory gauge coupling function is again 
captured by holomorphic sections of a certain line bundle. Such sections include a quadratic 
correction required for holomorphicity in the $T_\alpha$ coordinates, but also generally allow for 
as logarithm of a sum of Riemann theta functions with complex structure dependent coefficients. 
This line bundle and these theta functions are now defined 
on a complex torus spanned by the axions coming from the M-theory three-form that are not dualized
into vector multiplets in the F-theory frame. This torus is thus a subspace of the complex torus $H^{2,1}(Y_4)/H^3(Y_4,\bbZ)$,
which also captures the degrees of freedom of the R-R bulk vector fields. A detailed study of this 
geometric object and its variation over the complex structure moduli space is therefore of key phenomenological interest. 
In this work we have already conjectured certain
constrains on geometric data of elliptically fibered Calabi-Yau fourfolds. In particular, by demanding supersymmetry of the
four-dimensional effective action, we have proposed that the function $f_{\cA \cB}$, which is holomorphic in the complex structure moduli of $Y_4$ and defined in  \eqref{Psi_ansatz}, should satisfy some non-trivial relations. Our analysis has been done for a generic compactification space, without
referring to a specific example. Thus, it would be interesting to analyze in detail different examples to check whether such relations are
indeed satisfied.

Another interesting approach to derive the couplings relevant for the 
gauge coupling function in F-theory
was presented in a series of papers \cite{Lerche:1998nx,Lerche:1998gz,Lerche:1999hg}. 
It was shown in these papers that the coefficient functions of the couplings of type $F^4$, 
where $F$ is an eight-dimensional gauge field, satisfy certain Picard-Fuchs-type differential equations. 
It would be interesting to explore the relation of these findings to the results of this paper.

\subsection*{Acknowledgments}
		
We are grateful to Ralph Blumenhagen, Sebastian Greiner, Andreas Kapfer, Fernando Marchesano, Kepa Sousa, Stephan Stieberger, and Irene Valenzuela for useful discussions. This work was supported by a grant of the Max Planck Society.

\appendix

\section{Dualization of three-dimensional actions} \label{ap:dual}

In this appendix, we perform the dualization of $\mathcal{N}=2$ three-dimensional actions, where massless vectors and scalars are dual to each other. 
The dualization can be done explicitly by adding Lagrange multipliers to the action and integrating out the fields we want to dualize. In fact, this can be done in superspace, where the $\mathcal{N}=2$ supersymmetry is manifest, and which corresponds to a Legendre transform of the K\"ahler potential.

To illustrate this, let us consider the three-dimensional $\mathcal{N}=2$ action for a massless vector multiplet
\begin{equation}
\label{superspace_action}
S_1(V) = \int \dm 3 x \, \dm 2 \theta \, \dm 2 \widebar \theta \, \wt K(G(V))\,,
\end{equation}
where $G$ is the linear multiplet ($D^2G=\bar D^2G=0$) that contains the field strength, namely $G=\frac{i}{2} \bar D^\alpha D_\alpha V$.
To perform the duality transformation, we consider the parent action given by
\begin{equation}
S_P (G,\Phi)= \int \dm 3 x \, \dm 2 \theta \, \dm 2 \widebar \theta \, \left(\wt K(G) - G \re \Phi \right)\,,
\end{equation}
where $G$ is now an unconstrained real superfield and $\Phi$ is a chiral superfield. By varying $\Phi$, we find that $G$ is a linear superfield and substituting this in the action, we obtain \eqref{superspace_action}. On the other hand, varying the action with respect to $G$ gives
\begin{equation}
\re \Phi = \frac{\partial \wt K(G)}{\partial G}\,,
\end{equation}
which leads to the dual action
\begin{equation}
S_2(\Phi) = \int \dm 3 x \, \dm 2 \theta \, \dm 2 \widebar \theta \, K(\re \Phi) \,,
\end{equation} 
where $K$ is the Legendre transform of $\wt K$. 

Therefore, we may dualize the action used in the main text,
\begin{equation}
\label{action_start_A}
S^{(3)}_{\mathcal N=2} = \int - \wt K^{ \hat A {\bar{\hat B}}} \dd \phi_{{\hat A}} \ws \dd \wb  \phi_{{\bar{\hat B}}}  +   \frac 14 \wt K_{\S \L}( \dd L^\S \ws \dd L^\L + F^\S \ws F^\L) + F^\S \wedge \im (\wt K_{\S}^{\hat A}  \dd \phi_{\hat A})\,,
\end{equation}
by performing a Legendre transform of the kinetic potential. Since we want to dualize some of the scalars into vectors, and vice versa, we split the fields as follows\footnote{Notice that, for the dualization in section~\ref{Fthy}, we have that $\phi_a=(z^\cK,N_A)$.}
\begin{equation}
\phi_{\hat A} = (\phi_a,N_\k), \qquad L^\S = (L^\ih,L^\a), \qquad A^\S = (A^\ih, A^\a),
\end{equation}
and dualize the fields with Greek indices. We also assume that the kinetic potential does not depend on $\im N_\k$.\footnote{This is actually not strictly necessary, however, it is true for the K\"ahler potential that we dualize in the main text.} 
The appropriate Legendre transform is given by 
\begin{equation}
\label{legtransf}
K (\varphi_a, T_\a|l^\ih, n^\kappa) =\wt K(\phi_a,N_\k| L^\ih, L^\a) -  \,  L^\a \, \re T_\a  -  \, \re N_\k \, n^\k\,,
\end{equation}
where the new variables are defined as
\begin{equation}
\label{newvar2}
\re T_\a \equiv  \wt K_{ L^\a} \equiv \wt K_\a\,, \qquad n^\k \equiv  \wt K_{\re N_\k}\equiv \wt K^\k\,.
\end{equation}
The dual action takes exactly the same form as \eqref{action_start_A}, but with field content changed, 
\begin{equation}
\phi_{\hat A} = (\varphi_a,T_\a), \qquad L^\S = (l^\ih,n^\k), \qquad A^\S = (A^\ih, A^\k),
\end{equation}
and $K$ replaced by its Legendre transform given by \eqref{legtransf}. Although the fields $\phi_a$ and $L^\ih$ were not dualized, we nevertheless changed their names to $\varphi_a$ and $l^\ih$ for clarity. 

It is possible to express all the derivatives of $K$ in terms of those of $\wt K$, if one knows the derivatives of the old variables with respect to the new. The dualization gives us the opposite, i.e. the derivatives of the new variables with respect to the old, which we collect in a matrix
\begin{equation}
\label{Hessien}
M^i_j= \frac{\partial x_{\text{new}}^i}{\partial x_{\text{old}}^j} =  \ \begin{blockarray}{ccccl}
l^\ih &  \varphi_a &  \re T_\a & n^\k &\\[1.5mm]
\begin{block}{(cccc)@{\hspace{5mm}}l}
\delta^\ih_\jh & 0  &\wt K_{\a \jh} &\wt K^{\k}_{\jh} &  L^\jh       \\
0 & \delta_a^b & \wt K_{\a}^b &\wt K^{\k b} &          \phi_b   \\
0 & 0  & \wt K_{\a \b} & 0 &      L^\b     \\
0 & 0  &0 &\wt K^{\k \lambda} &                \re N_\lambda\\
\end{block}      
\end{blockarray}
\arraycolsep=1mm
\equiv \ \begin{pmatrix}
\, \mathds{1} &  B \, \\
\, 0 &  D \,
\end{pmatrix} \,,
\end{equation}
where we assumed that $\wt K_\a^\k=0$.\footnote{This is true for the K\"ahler potential \eqref{originalK}, since $d_\a{}^{\k \cA} = 0$. It is straightforward to drop this assumption.}
The derivatives of the old variables in terms of the new ones is given by the inverse of this matrix, namely
\begin{equation}
\label{Minv}
M^{-1} = 
\begin{pmatrix}
\, \mathds{1} & - B D^{-1} \\
\, 0 &  D^{-1}
\end{pmatrix}
= \begin{blockarray}{ccccl}
L^\ih & \phi_a  & L^\a & \re N_\k & \\[1.5mm]
\begin{block}{(cccc)@{\hspace{5mm}}l}
\delta^{\ih}_\jh & 0& -\wt K_{\b \jh} (\wt K_{\a \b})^{-1} & -\wt K^{\lambda}_{\jh} (\wt K^{\k \lambda})^{-1} &  l^\jh       \\[1mm]
0 & \delta_{a}^b & -\wt K_{\b}^{b} (\wt K_{\a \b})^{-1} & - \wt K^{\lambda b}(\wt K^{\k \lambda})^{-1} &         \varphi_b   \\[1mm]
0 & 0 &  (\wt K_{\a \b})^{-1} & 0 &          \re T_\b        \\
0 & 0 & 0 & (\wt K^{\k \lambda})^{-1} &   n^\lambda\\
\end{block}      
\end{blockarray} \,.
\end{equation}
Using this we find the derivatives of the new kinetic potential in terms of derivatives of the original one, which are given by
\begin{equation}
\begin{alignedat}{4}
& K_{\k \lambda} &&=  -  (\wt K^{\k \lambda})^{-1}      &&  K_{\k i} &&=\wt K_i ^\lambda (\wt K^{\k \lambda})^{-1}      \\
& K^{ \a\bar{\b}} &&=  - \mfrac 14 (\wt K_{\a \b})^{-1}     &&  K^{ a}_{\k} &&=\wt K^{\lambda a}(\wt K^{\k \lambda})^{-1}   \\
& K_{i j} &&=\wt K_{ij}   -\wt K_i ^\k\wt K_j^\lambda (\wt K^{\k \lambda})^{-1} - \wt K_{i \a}\wt K_{j \b} (\wt K_{\a \b})^{-1}          &&  K^{ \a a}  &&= \mfrac 12\wt  K_\b^a (\wt K_{\a \b})^{-1}                \\
& K^{a}_i &&=  \wt K^a_{i}   - \wt K_i^\k\wt K^{a \lambda}(\wt K^{\k \lambda})^{-1}   - \wt K_{i \a} \wt K^a_{\b}(\wt K_{\a \b})^{-1} &&  K^{ \a}_{i} &&= \mfrac 12\wt K_{\b i} (\wt K_{\a \b})^{-1}       \\
& K^{ ab} &&=\wt K^{ab}-\wt K^{a \k }\wt K^{\lambda b}(\wt K^{\k \lambda})^{-1}  - \wt K_\a^a\wt K^b_{\b} (\wt K_{\a \b})^{-1}  
\qquad \quad  &&   K^{ \a}_{\k} & &= 0 \,.              
\end{alignedat}
\end{equation}
For the case analyzed in the main text, namely for $\wt K$ given in \eqref{originalK}, we find the following derivatives of $K$
\begin{align}\label{app1}
 K_{\k\l} = & \, \,  \frac 1R \re f_{\k \l}\\
  K_{\k i} = & \, \, \frac 1R \re [Q_{i\k}{}^AN_A] \\
 K_{\k \cK} = & \, \, \frac{1}{2R} \, \partial_\cK f_{\k\l} (n^\l +i L^i M_i{}^{\l A} N_A) = \frac{1}{2R} \left ( \partial_{\cK} f_{\k\l} \, n^\l + \partial_\cK Q_{ik}{}^A \,L^i N_A \right ) \\ K^{A}_{\k} = & \, \, \frac 1{2R} Q_{i\k}{}^A \, L^i \\
 K^{A}_i = & \, \,  \frac 1{2R} Q_{i\k}{}^{A} n^\k - \frac{1}{R} \left(Q_{i\k}{}^A \re [d_j{}^{B\k}N_B] + Q_{j\k}{}^A \re[d_i{}^{B\k}N_B] \right)\, L^j \notag \\ & +\frac{1}{2R} \left( \re d_\a{}^{AB}N_B +d_\a{}^{AB}\wb N_B \right) C^\a_{ij} L^j \\
 K^{}_{i\cK} = & \, \, \frac 1R \, \partial_\cK f_{\k\l} \Big[- 2 \left( \re [d_j{}^{A\k}N_A]d_i{}^{\lambda B}\re N_B +i \re[d_i{}^{A\k}N_A]\im d_j{}^{\lambda B} N_B\right) L^j  \notag \\
 & +  n^\k i\im d_i{}^{\l A} N_A \Big ] - \frac{1}{2R}\, \partial_\cK f_{BC}  \left(\re f^{AB}  \, d_\a{}^{CD} \re N_A \re N_D \right) C^\a_{ij} L^j \\
K_{ij} 
= & \ \frac 1R \re \left[ -C_{ij}^\a \left( T_\a - \re[d_\a{}^{AB}N_B]\re N_A\right)
+4 \re[d_i{}^{A\k}N_A]\re [d_j{}^{B\lambda}N_B] \re f_{\k\lambda} \right]\\
K^{ \a}_i = & \, \, - \frac 1{2R} C_{ij}^\a L ^j  \\
K^{ \a\bar{\b}} = & \, \, \frac 1{16} (G_{\a\b})^{-1} + \frac 1{16 R } (G_{\a\gamma})^{-1} (G_{\b\delta})^{-1} H_{\gamma \delta\epsilon} C^\epsilon_{ij} L^i L^j   \,,  
\end{align}
where we defined 
\begin{align}
G_{\a\b} &\equiv -\frac 32 \left(\frac{\hK_{\a\b}}{\hK} - \frac 32 \frac{ \hK_\a \hK_\b}{(\hK)^2} \right)\\
H_{\a\b \gamma} &\equiv - \frac 34 \left( \frac{\hK_{\a\b\gamma}}{\hK}  - 3 \frac{\hK_{\a\b} \hK_\gamma +\hK_{\b\gamma}\hK_\a +\hK_{\gamma\a}\hK_\b}{(\hK)^2}  + 9 \frac{\hK_\a \hK_\b \hK_\gamma}{(\hK)^3}   \right) \,,
\end{align}
and worked at leading order in $C_{ij}^\a$. 
It is also useful to consider the following combinations
\begin{dgroup*}
	\begin{dmath}
		F^\k \wedge \im (K^{ \hat A}_\k \dd \phi_{\hat A}) = \frac{n^\lambda}{2R} F^\k \wedge \dd \im f_{\k\l} + \frac{L^i}{2R} F^\k \wedge \dd \im ( Q_{i\k}{}^A N_A) 
	\end{dmath} 
	\begin{dmath} 
		F^i \wedge \im (K_i^{ \hat A}\dd \phi_{\hat A}) 
		= \frac{n^\k}{2R} F^i \hiderel{\wedge} \dd \im \left( Q_{i\k}{}^A N_A \right) 
		+  \frac{L^j}{R} F^i \hiderel{\wedge} \dd\im \left[-C_{ij}^\a \left(\mfrac 12 T_\a - \re[d_\a{}^{BA}N_B]\re N_A\right) 
		+4 \re[d_i{}^{A\k}N_A]\re [d_j{}^{B\lambda}N_B] \re f_{\k\lambda}\right]\,. \hspace*{7mm}
	\end{dmath}
\end{dgroup*}

\section{Circle reduction of four-dimensional $\mathcal N=1$ supergravity} \label{ap:circle}

In this appendix, we perform the circle reduction of the following four-dimensional $\mathcal{N}=1$ ungauged supergravity action,
\begin{equation}
\begin{aligned}
S ^{(4)} = \int_{\mathcal{M}_4}\frac 12  \hat R \, \hat \star \, 1 - \hat K_{\bA \wb \bB} \dd \hat M^\bA \wedge \hat \star \dd \widebar{\hat M}^{\bar \bB} &- \frac 14 \re f_{\bI \bJ} (\hat M) \hat F^\bI \wedge \hat \star \hat F^\bJ \\
&\hspace{1cm }- \frac 14 \im f_{\bI \bJ} (\hat M ) \hat F^\bI \wedge \hat F^\bJ   \, ,
\end{aligned}
\end{equation}
where $\hat K_{\bA\wb \bB}$ and $\re f_{\bI \bJ}$ are positive definite (we use the mostly minus metric convention), and hatted objects live in four dimensions. 
When $\mathcal{M}_{4}=\mathcal{M}_3 \times S^1$, we can decompose the metric as
\begin{equation}
\label{metric}
\dd \hat s^2 = \dd s^2 + r^2 \, (\dd y +A^0)^2, \qquad y \sim y +2\pi\, . 
\end{equation}
With such a decomposition of the metric, one finds the following reduction of the Einstein-Hilbert term
\begin{equation}
		\int_{\mathcal{M}_4}  \frac 12   R_4 \,\hat{\star}\, 1 
		=  \int_{\mathcal{M}_3}   \frac 12  R_3 \star 1 - \frac{1}{4R^2} \dd R \wedge \star \dd R - \frac 1{4R^2} F^0 \wedge \star F^0 \, ,
		\label{gravpart}
\end{equation}
where in addition, we performed a Weyl rescaling $g_{\mu \nu}^{\mathrm{new}} = r^2 g_{\mu \nu}^{\mathrm{old}}$ to bring the action to the Einstein frame, and introduced the new variable $R\equiv r^{-2}$.

Furthermore, the reduction ansatz for the vectors is,
\begin{equation}
\hat A^\bI = A^\bI - \zeta^\bI \, (\dd y + A^0) \, ,
\end{equation}
where $\zeta^\bI$ are three-dimensional scalars. The reduction of the terms containing vectors is
\begin{dgroup*}
	\begin{dmath}
		\int_{\mathcal{M}_4}\! \re f_{\bI\bJ} \, \hat F^\bI \wedge \star\, \hat F^\bJ = \int_{\mathcal{M}_3} \frac 1R \re f_{\bI\bJ} \left [ \left (F^\bI - \frac{\xi^\bI}{R} F^0 \right ) \wedge \star \left (F^\bJ - \frac{\xi^\bJ}{R} F^0 \right ) \hspace*{13mm} \\ 
				\hspace*{2.4cm}- \left( \dd \xi^\bI -\frac{\xi^\bI}{R} \dd R  \right) \wedge \star \left ( \dd \xi^\bJ - \frac{\xi^\bJ}{R} \dd R \right ) \right ]\,,
				\label{realpart}
	\end{dmath}
	\begin{dmath}
	\int_{\mathcal{M}_4}\im f_{\bI\bJ} \, \hat F^\bI \wedge \hat F^\bJ 
		=  2 \int_{\mathcal{M}_3} \dd \im f_{\bI\bJ} \wedge \left (F^\bI - \frac 12 \frac{\xi^\bI}{R} F^0 \right ) \frac{\xi^\bJ}{R}\,,
		\label{impart}
	\end{dmath}
\end{dgroup*}
where we introduced $\displaystyle \xi^\bI \equiv R\,\zeta^\bI$, which are the proper three-dimensional scalar fields (they form a vector multiplet together with the reduced vector $A^\bI$ ; similarly $R$ and $A^0$ form a vector multiplet).

Putting all this together we obtain the following three-dimensional action
\begin{equation}
\begin{aligned}
	S^{(3)}  = \int_{\mathcal{M}_3} \ & \frac 12 R_3 \star 1 - \frac{1}{4R^2}\left ( \dd R \ws \dd R + F^0 \ws F^0 \right ) 
	- \hat K_{\bA\bar \bB} \dd M^\bA \wedge \star \dd \bar M^{ \bB } \\ 
	& - \, \frac{1}{4R} \re f_{\bI\bJ} \left( \dd \xi^\bI -\frac{\xi^\bI}{R} \dd R  \right) \ws \left ( \dd \xi^\bJ -\frac{\xi^\bJ}{R} \dd R \right )\\
	&-\frac{1}{4R}\re f_{\bI\bJ}  \left (F^\bI {-} \frac{\xi^\bI}{R} F^0 \right ) \ws \left (F^\bJ {-} \frac{\xi^\bJ}{R} F^0 \right ) \\
	& - \frac 12  \dd \im f_{\bI\bJ} \wedge \left (F^\bI  {-} \frac 12 \frac{\xi^\bI}{R} F^0 \right ) \frac{\xi^\bJ}{R} \, .
	\label{3dactionexp}
\end{aligned}
\end{equation}
One can check that this action can be put into the standard $\mathcal{N}=2$ supergravity form,
\begin{equation}
\begin{aligned}
\label{act:3D}
S^{(3)}&=\int_{\mathcal{M}_3} \frac 12 R_3 \star 1 -  K_{\bA\bar \bB} \dd M^\bA \wedge \star \dd \widebar M^{ \bB } \\ 
&\hspace*{2cm} + \frac 14 {K}_{\cI \cJ} \left(\dd \xi^{\cI} \wedge \star \dd \xi^{\cJ} + F^{\cI} \wedge \star F^{\cJ}\right) +  F^{\cI} \wedge \im({K}_{\cI \bA}\dd M^\bA) \, ,
\end{aligned} 
\end{equation}
with kinetic potential 
\begin{equation*}
{K} = \hat K(M,\widebar M) + \log R - \frac{1}{2R} \re f_{\bI \bJ} \, \xi^\bI \xi^\bJ \, , 
\end{equation*}
where the indices $(0,\bI)$ have been gathered into a single index $\cI$,
\begin{equation*}
\xi^{\cI} = (R,\xi^\bI), \quad A^{\cI} = (A^0,A^\bI)\, . \label{defs}
\end{equation*} 

\bibliographystyle{utcaps}

\providecommand{\href}[2]{#2}\begingroup\raggedright

\end{document}